\documentclass[12pt,thmsb,titlepage]{article}
\usepackage{amsmath}
\usepackage{amssymb}
\usepackage{graphicx}
\usepackage{epsf}
\usepackage[T1]{fontenc}
\usepackage[latin9]{inputenc}
\usepackage[letterpaper]{geometry}
\usepackage{setspace}
\usepackage{epstopdf}

\usepackage{rotating}

\usepackage{epstopdf}
\usepackage{epsf}

\newcommand{\captionfonts}{\footnotesize}
\makeatletter  
\long\def\@makecaption#1#2{%
  \vskip\abovecaptionskip
  \sbox\@tempboxa{{\captionfonts #1: #2}}%
  \ifdim \wd\@tempboxa >\hsize
    {\captionfonts #1: #2\par}
  \else
    \hbox to\hsize{\hfil\box\@tempboxa\hfil}%
  \fi
  \vskip\belowcaptionskip}
\makeatother   

\newenvironment{shrinkeq}[1]
{ \bgroup
  \addtolength\abovedisplayshortskip{#1}
  \addtolength\abovedisplayskip{#1}
  \addtolength\belowdisplayshortskip{#1}
  \addtolength\belowdisplayskip{#1}}
{\egroup}

\newtheorem{theorem}{Theorem}

\newtheorem{lemma}{Lemma}
\newtheorem{corollary}{Corollary}

\begin{document}

\centerline{\Large{\bf Factor models with many assets: }}
\centerline{\Large{\bf strong factors, weak factors, and the two-pass procedure}}
\medskip

\centerline{by Stanislav Anatolyev\footnote{CERGE-EI and New Economic School. Address: Politick\'{y}ch v\v{e}z\v{n}\r{u} 7, 11121 Prague, Czech Republic. E-mail: stanislav.anatolyev@cerge-ei.cz. Czech Science Foundation support under grant number 17-26535S is gratefully acknowledged.}
and Anna Mikusheva\footnote{
Department of Economics, M.I.T. Address: 77 Massachusetts Avenue, E52-526, Cambridge, MA, 02139. Email: amikushe@mit.edu. National Science Foundation support under grant number 1757199 is gratefully acknowledged. We thank David Hughes for research assistance.}}
\medskip

\centerline{\bf Abstract}
\smallskip

\noindent {\small{This paper re-examines the problem of estimating risk premia in unconditional linear factor pricing models. Typically, the data used in the empirical literature are characterized by weakness of some pricing factors, strong cross-sectional dependence in the errors, and (moderately) high cross-sectional dimensionality. Using an asymptotic framework where the number of assets/portfolios grows with the time span of the data while the risk exposures of weak factors are local-to-zero, we show that the conventional two-pass estimation procedure delivers inconsistent estimates of the risk premia. We propose a new estimation procedure based on sample-splitting instrumental variables regression. The proposed estimator of risk premia is robust to weak included factors and to the presence of strong unaccounted cross-sectional error dependence. We derive the many-asset weak factor asymptotic distribution of the proposed estimator, show how to construct its standard errors, verify its performance in simulations, and revisit some empirical studies.
}
\smallskip

\noindent\texttt{Key words:} factor models, price of risk, risk premia,
two-pass procedure, strong factors, weak factors, dimensionality
asymptotics, weak factor asymptotics.
\smallskip

\noindent\texttt{JEL classification codes:} C33, C38, C58, G12.}

\section{Introduction}

\label{section-introduction}

Since the introduction of the CAPM by Sharpe (1964) and Linner (1965), linear
factor pricing models have grown into a very popular sub-field in asset
pricing. Harvey, Liu and Zhu (2016) list hundreds of papers that propose,
justify and estimate various factor pricing models. A typical paper in
this area proposes a small set of observed risk
factors that price the assets, that is, the expected excess return on an asset is equal to the quantity of risk taken (measured as a normalized covariance of the returns with the risk factors, so called betas) times risk premia.  The two most famous factor pricing models are the market-factor CAPM and the three-factor Fama and French (1993) model. Other pricing factors are  the momentum factor (Jegadeesh and Titman, 1993), the consumption-to-wealth ratio `cay' (Lettau and Ludvigson, 2001), the liquidity factor (P\'{a}stor and Stambaugh, 2003), and so on. In recent years, there has been a burst in econometrics research that suggests how to correct the baseline estimation and inference in the face of
many factors (for example, Kozak, Nagel, and Santosh, 2018), or how to judiciously select factors from a big pool without jeopardizing correct inference (for example, Feng, Giglio and Xiu, 2019).

Traditionally, one estimates the model using what is commonly known as the
two-pass estimation procedure (Fama and MacBeth, 1973; Shanken, 1992),\footnote{%
Sometimes the two-pass procedure is referred to as the Fama-MacBeth procedure
(Fama and MacBeth, 1973). See Cochrane (2001, section 12.3) on their
numerical equivalence when betas are time invariant. The method for obtaining
valid standard errors that account for the two step nature of the procedure
is given in Shanken (1992).} where at the first pass one estimates risk exposures (betas) for each asset, and then, at the second pass, those estimates are used as regressors to estimate the risk premia. Asymptotic justification of this procedure, however, relies on  assumptions that often do not hold up in realistic circumstances. Two types of violations of the idealistic setting have been noted in  previous literature.

The first problem is one of weak (but priced) observed factors. Recent papers by Kan and Zhang (1999), Kleibergen (2009), Bryzgalova (2016), Burnside (2016), and Gospodinov, Kan and
Robotti (2017) all point out that risk exposures (or betas) to some observed
factors tend to be small to such an extent that their estimation errors are
of the same order of magnitude as the betas themselves. This observed phenomenon
is very similar to the  widely studied weak instrument problem.

The second violation is a strong cross-sectional dependence in error terms, which in many cases can be modeled as a factor structure unaccounted for (`missing'). For example, recent literature shows that mismeasurement of the true risk factors leads to  weakness of the observed factors and strong cross-sectional dependence in the errors (Kleibergen and Zhan, 2015), which
may result in all sorts of distortions in estimation and inference in
theory and in their non-reliability in practice (Kan and Zhang, 1999;
Andrews, 2005; Kleibergen, 2009). 

Along with the combination of the problems of  missing factors and small betas, we also
consider one very important empirical feature of the typically employed datasets
-- the presence of  a large number of assets or portfolios often comparable to the number of
periods over which returns are observed. We consider an asymptotic framework
where the number of assets/portfolios grows with its time-series dimension. Such dimension
asymptotics is likely to provide a more accurate asymptotic approximation to the
finite sample properties of estimators and tests.  The many-asset
asymptotic framework has been utilized previously by Gagliardini,
Ossola and Scaillet (2016a), Lettau and Pelger (2018) and Feng, Giglio and Xiu (2019).

We show that within a dimension-asymptotic framework the presence of small
betas leads to a failure of the classical two-pass procedure, while the
additional presence of missing factors exacerbates this problem. We propose
econometric procedures that are robust to  both these thorny issues with factors --
the weakness of observed factors and the presence of unobserved factors in
the errors -- and, in contrast to the remedies proposed elsewhere, are
easily implementable using standard regression tools (in particular,
instrumental variables regressions and two-stage least squares). The
estimators we propose are consistent and asymptotically mixed gaussian; moreover,
using the variance estimators (the construction of which we describe),  standard
inference tools such as $t$- and Wald tests can be applied in a conventional
way.

Our new estimation approach makes use of the idea of sample-splitting in order to create  multiple estimates for loadings $\beta_i$ and to correct for the first-step estimation error via  an instrumental variables regression. The presence of an unobserved (missing) factor structure in the error terms  creates strong cross-sectional dependence in the panel of returns, which is similar to the classical omitted-variables problem in the second pass of the
two-pass procedure. In order to correct for this missing factor structure, we use sample splitting to create reasonable proxies for missing factors even in a setting where one cannot consistently estimate the missing factor structure. The sample-splitting idea has appeared in the econometrics literature before, in particular, in Angrist and Krueger (1992) and Dufour and Jasiak (2001).

We explore the quality of the two-pass procedure and compare its performance with that of sample-splitting based estimators in simulations calibrated to match the monthly returns of the 100 Fama-French sorted portfolios. We applied both procedures to the estimation of the momentum risk premium using real data on Fama-French portfolios. The important feature here is that the momentum is a tradable factor and hence there is an alternative estimate of the risk premia -- the sample average excess return on this factor. Thus, we have a natural benchmark when comparing two estimators, which differ greatly. Lewellen, Nagel and Shanken (2010) showed that
``any (sufficiently large)\footnote{Our addition in brackets.} set of assets perfectly explains the cross-section of expected returns so long as the  (tested)\footnote{Our addition in brackets.} assets are not asked to price
themselves (that is, ...the risk premia are not required to equal their expected returns).'' From that perspective, having an estimate of risk premia coming from the pricing model, such as our split-sample estimator, is important even for tradable factors, as it allows one to test the pricing model by comparing this estimate to the average excess return.

There is a growing number of alternative suggestions for how to correct statistical inferences for either weak observed factors or a missing factor structure, but we are not aware of any paper that solves both problems simultaneously. Kleibergen (2009) proposes the use of weak identification robust inference procedures to account for weak observed factors, however, it can only be applied to a relatively small number of assets/portfolios. An extreme version of the weak factors phenomenon, known as irrelevant factors and studied by Bryzgalova (2016), Burnside (2016) and  Gospodinov, Kan and Robotti (2017), occurs when some observed factors are assumed to have zero loadings. The solutions to the irrelevant factors problem proposed in the literature usually suggest dimension reduction techniques to  detect the irrelevant factors with a proviso to eliminate these detected irrelevant factors from further analysis.  However, applying detection and elimination methods to the weak observed factors would lead to invalid inferences and large biases in the estimates of the risk premia for the remaining factors.
Jegadeesh, Noh, Pukthuanthong, Roll and Wang (2019) suggest, simultaneously and independently, the use of sample-splitting in factor models in order to fix the errors-in-variables bias. Their proposed estimator works only when there are no missing factors in the error terms.

Giglio and Xiu (2017)  solve the problem of the missing factor structure by first running the Principle Component Analysis (PCA) on excess returns, pricing principle components, and from that deriving the risk premia of observed factors. The method of Giglio and Xiu (2017) successfully eliminates  strong missing factors, but assumes from the outset that all important pricing factors can be uncovered by PCA -- essentially, they assume away any existence of weak factors needed for pricing. This assumption is critical for the validity of their procedure and contradicts the empirical findings of Lettau and Pelger (2018), who demonstrate that the out-of-sample performance using  weak factors in addition to those uncovered by PCA is sizably better than that of a model that uses the PCA factors only. According to Lettau and Pelger (2018), ``PCA-based factors often miss low volatility components with high Sharpe ratios, which is a crucial aspect in asset pricing.''

The paper is organized as follows. Section \ref{section-setting} introduces
notation, discusses the relevance of our asymptotic approach, and argues for the
presence of a significant factor structure in the errors. Section \ref{section-
assumptions} introduces and discusses technical assumptions. Section \ref%
{section- FMB fail} explains the asymptotic failure of the classical
two-pass procedure and provides detailed intuition as to why this occurs. We propose our `four-split' estimation method in Section \ref{section- our method}, describe what
motivates it  and explain why it works. We also state a formal theorem on the
consistency of the newly-proposed four-split estimator. Section \ref{section- inference} is devoted to deriving inference procedures that use our four-split estimator; in particular, we show the asymptotic validity of a properly constructed Wald test. The size of a potential failure of the two-pass procedure and the behavior of the newly proposed estimator are studied in simulations and with an empirical example in Section \ref{section- simulations}. Some proofs and additional results appear in a Supplementary Appendix available on the authors' web-sites.\footnote{\texttt{http://economics.mit.edu/files/15225}}

A word on notation: $0_{l,m}$ stands for a zero matrix of size $l\times m$, $%
I_{m}$ is an $m\times m$ identity matrix; for an $m\times l_{A}$ matrix $A$ and an $%
m\times l_{B}$ matrix $B$, $(A,B)$ stands for the $m\times (l_{A}+l_{B})$
matrix one obtains by placing the initial matrices side-to-side. For a square matrix $A$ we denote $\mathrm{dg}(A)$ a diagonal matrix of the same size with the same elements on the diagonal as matrix $A$.

\section{Formulation of the problem}

\label{section-setting}

The research in factor asset pricing modeling typically proposes a small set of observed risk factors described by a vector $F_{t}$ of (usually) small dimension $k_{F}$. An
asset or portfolio of assets $i$ with excess return $r_{it}$ has exposure to
several risk factors, which is quantified by the asset's betas $\beta
_{i}=\mathrm{var}(F_{t})^{-1}\mathrm{cov}(F_{t},r_{it})$.  A typical claim put forth in the linear factor-pricing theory is that exposure to risk (betas) fully determines the assets'
expected excess returns. Particularly, there exists a $k_{F}$-dimensional
vector of risk premia $\lambda $ such that $Er_{it}=\lambda ^{\prime }\beta
_{i}$. This is known as unconditional factor pricing setting, and some generalizations to conditional factor pricing will be discussed later on.

From an econometric perspective, a correctly-specified linear factor-pricing model is equivalent to the following formulation:
\begin{shrinkeq}{-1ex}
\begin{equation}  \label{eq: initial FMB model}
r_{it}=\lambda ^{\prime }\beta _{i}+(F_{t}-EF_{t})^{\prime }\beta
_{i}+\varepsilon _{it},
\end{equation}
\end{shrinkeq}%
where unobserved random error terms $\varepsilon _{it}$ have mean zero and are
uncorrelated with $F_{t}$. Here the statement $E\varepsilon_{it}=0$ is equivalent to $Er_{it}=\lambda ^{\prime }\beta
_{i}$, while uncorrelatedness between $\varepsilon_{it}$ and $F_t$ results from the definition of $\beta_i$. We treat $\lambda$ and $\beta_i$ as unknown parameters, while $r_{it}, F_t,$ and $ \varepsilon_{it}$ are random variables. One special case often mentioned in the
literature occurs when the $F_{t}$ factors  are asset returns themselves and are
supposed to be priced by the same model; in this case, theoretically $%
\lambda =EF_{t}$. We will not make this assumption and will consider the
general case when $\lambda $ differs from $EF_{t}$.

\textbf{Two-pass procedure.}
The estimation and  inferences on risk prices, $\lambda $, are
usually accomplished by a procedure commonly known as the two-pass
procedure (Fama and MacBeth, 1973; Shanken, 1992), which is
applied to a data set consisting of a panel of asset excess returns $%
\{r_{it},$ $i=1,...,N,$ $t=1,...,T\}$ and  observations of realized factors $%
\{F_{t},$ $t=1,...,T\}$. In the first step, one estimates $\beta _{i}$ by
running a time series OLS regression of $r_{it}$ on a constant and $F_{t}$
for each $i=1,...,N$. The second step produces an estimate of $\lambda $
(denote it $\widehat{\lambda }_{TP}$) by regressing the time-average excess
return $\frac{1}{T}\sum_{t=1}^{T}r_{it}$ on the first-step estimates, $%
\widehat{\beta }_{i}$. Under suitable conditions, $\widehat{\lambda }_{TP}$
is proved to be both consistent and asymptotically gaussian. Discussions of
the statistical properties of the two-pass procedure appear in Fama
and MacBeth (1973), Shanken (1992), and Chapter 12 of Cochrane (2001).

This paper deviates from the classical Fama-MacBeth setting in three respects, which we label as (i) weak observed factors, (ii) many assets and (iii) missing factor structure.

\emph{Weak observed factors.} Recent work by several prominent researchers  raises the concern that the
two-pass procedure may provide misleading estimates of risk premia; see, for
example, Kan and Zhang (1999), Kleibergen (2009), Bryzgalova (2016),
Burnside (2016), Gospodinov, Kan and Robotti (2017). They surmise that the
reason for these erroneous inferences is attached to the empirical observation
that either some column of $\beta =(\beta _{1},...,\beta _{N})^{\prime }$ is
close to zero, or, more generally, the $N\times k_{F}$ matrix $\beta $
appears close to one of reduced rank (less than $k_{F}$) for many
well-known linear factor pricing models.
According to Lettau and Pelger (2018), weak factors are empirically crucial for good performance of pricing models.   They constructed  factors that are impossible to uncover by PCA (i.e. weak factors) with the Sharpe ratio twice as high as those uncovered by PCA, and the out-of-sample pricing errors from a model that uses these weak factors are sizably smaller than those a the model that uses only strong factors.

Bryzgalova (2016), Burnside (2016), and Gospodinov, Kan and
Robotti (2017) all recently developed improved  inference procedures when some factors are completely \emph{irrelevant} for pricing, that is, when true $\beta_i$ are exactly zeros. The above-mentioned papers propose strategies for detecting and eliminating irrelevant factors, which is important as the risk premia on the irrelevant factor are completely unidentified. Unfortunately, these procedures  fail when the $\beta_i$ are not zeros, but are rather small.

A more empirically relevant case, which is in line with Lettau and Pelger (2018), resembles
 the widely studied weak instrument problem (Staiger and Stock, 1998): if
some of the observed factors $F_{t}$ are only weakly correlated with all
the returns in the data set, then the noise that arises in the first-pass
estimates of the corresponding components of $\beta _{i}$ will dominate the
signal, and the second-pass estimate of the risk premia $\lambda $ will be
oversensitive to small perturbations in the sample.
In order to model the observed phenomenon, Kleibergen (2009) considered a
drifting-parameter framework in which some component of $\beta
_{i} $ is modeled to be of order $O(\frac{1}{\sqrt{T}})$ assuming that the number of
time periods, $T$, increases to infinity, while the number of assets, $N$,
stays fixed. In such a setting the first-pass estimation error is of order
of magnitude $O_{p}(\frac{1}{\sqrt{T}})$, which is comparable to the size of
the coefficients themselves. This framework implies inconsistency
of the two-pass estimates for the risk premium on small components, poor
coverage of regular confidence sets even for the risk premium of strong
factors, and asymptotic invalidity of classical specification tests and
tests about risk premia.

Following this tradition and acknowledging the empirical evidence provided in
Kleibergen and Zhan (2015) and Bryzgalova (2016), we also make use of
drifting-parameter modeling. We assume that the $k_{F}\times 1$ vector of
factors $F_{t}$ can be divided into two subvectors: a $k_{1}\times 1$
dimensional vector $F_{t,1}$ and a $k_{2}\times 1$ vector $F_{2,t}$ (here $%
k_{F}=k_{1}+k_{2}$) such that the risk exposure $\beta _{i,1}$ to factor $%
F_{t,1}$ will be strong, while the risk exposure coefficients $\beta _{2,i}$
to factor $F_{2,t}$ will be drifting to zero at the rate $\sqrt{T}$.
We make these order assumptions for risk exposures more accurate in the next
section. The most important feature of our modeling is that the
standard error of the first-pass estimator of $\beta _{2,i}$ will be of the
same order of magnitude as the coefficient itself. A more general treatment
of the near-degenerate rank condition considers some $k_{2}$-dimensional
linear combination of factors (unknown to the researcher) to have a
local-to-zero (of order $O(\frac{1}{\sqrt{T}})$) exposure coefficient, while
the exposure to risk formed by the orthogonal $k_{1}$-dimensional linear
combination remains fixed. All our results are easily generalizable to this
setting, as we do not assume that the researcher knows which factors (or
combination of factors) bear small coefficients of exposure. However, to
simplify the exposition we will stick to the division of factors into two
sub-vectors.

\emph{Many assets.} In theoretical justifications of the two-step procedure (Shanken, 1992; Cochrane, 2001) it is common to assume that the number of assets, $N$, is fixed, while the number of periods $T$ grows to infinity.
We notice that in many common data sets that researchers use, the number of assets is large when compared to the
number of time periods. The celebrated Fama-French data set provides returns
on $N=25$ sorted portfolios for about $T=200$ periods. The often-used
Jagannathan-Wang data set (Jagannathan and Wang, 1996) contains
observations on $N=100$ portfolios observed for $T=330$ periods. Lettau and
Ludvigson (2001) use Fama-French $N=25$ portfolios, the returns for which
are observed over $T=141$ quarters. Gagliardini, Ossola and Scaillet
(2016a) use $N=44$ industry portfolios observed during $T=546$ months. In
these cases it is hard to believe that the asymptotic results derived under
the assumption that $N$ is fixed would provide an accurate approximation of
finite-sample distributions. Indeed, among other things, Kleibergen (2009)
discovers that the bias of the two-pass estimate of risk premia is strongly
and positively related to the number of assets if the total factor strength
is kept constant.

In this paper we consider asymptotics when both $N$ and $T$ increase to
infinity without restricting the relative growth between them.
Two recent papers by Raponi, Robotti and Zaffaroni (2017) and Kim and Skoulakis (2018) consider a factor pricing model in an asymptotic setting with $N\to \infty$ while $T$ remains fixed. They show an inconsistency of the two-pass procedures, which arises from the fact that with fixed $T$ loadings $\beta_i$ are estimated from a finite sample; as a result, the estimation error of $\beta_i$ is non-negligible. This is exactly the same phenomenon we are modeling with weak factors when we consider $T\to\infty$ -- namely, non-negligibility of the first step estimation errors. We can show that the procedures we propose are consistent in the setting with $N\to\infty$, fixed $T$ for \emph{ex-post risk premia} (see Shanken, 1992, for distinction with ex-ante risk premia)  if we impose slightly stronger assumptions on the cross-sectional dependence of the error terms than the one introduced in Assumptions ERRORS below.

\emph{Missing factor structure.} This paper's main deviation  from the existing literature is
our explicit acknowledgment of high cross-sectional dependence among error
terms $\varepsilon _{it}$ in model (\ref{eq: initial FMB model}). In
particular, we assume that errors have a factor structure. Namely, this
means that there exists, unknown and unobserved to the researcher, a
factor $v_{t}$ and loadings $\mu _{i}$ such that
\begin{equation*}
\varepsilon _{it}=v_{t}^{\prime }\mu _{i}+e_{it},
\end{equation*}
where the `clean' errors $e_{it}$ are only weakly cross-sectionally
dependent to the extent that asymptotically we may ignore their
dependence (the exact formulation of this assumption  appears in the next
section). Similar weak-dependence assumptions appear in approximate factor
models (e.g., Bai and Ng, 2002). The assumptions on loadings $\mu _{i}$
 guarantee that the factor structure will be strong enough to be both detected
empirically and asymptotically important for inferences. An insightful
discussion of weak vs strong factor structure and cross-sectional
dependence appears in Onatski (2012).

Below we provide two theoretical reasons as to why we expect to encounter factor structure in many linear factor-pricing models. Then we point to empirical
evidence that a missing factor structure is indeed present in some
well-known factor-pricing models.

\textbf{Example 1.} This example is similar to the one discussed in Lewellen, Nagel and Shanken (2010).
If one does not observe the true risk factors that price assets but only
proxies for them, this would lead to a factor structure in errors (see
Kleibergen and Zhan, 2015). Assume for a moment that
the market is priced by risk premia on risk factors $G_{t}$. For
expositional simplicity we assume that
\begin{shrinkeq}{-2ex}
\begin{equation*}
r_{it}=G_{t}\beta _{i}^{G}+\varepsilon _{it}^{G},
\end{equation*}
\end{shrinkeq}%
where the shocks $\varepsilon _{it}^{G}$ are drawn with mean zero and finite
variance independently cross-sectionally, and independent
from $G_{t}$. Assume that $G_{t}$ is stationary with variance $\Sigma
_{G}$.

Assume that the econometrician does not observe $G_{t}$, but rather has a
proxy for it, $F_{t}=\alpha +\delta G_{t}+\epsilon _{t}$, where $\epsilon
_{t}$ has mean zero and is uncorrelated with $G_{t}$ and shocks $\varepsilon
_{it}^{G}$. For example, $\epsilon _{t}$ may stand for a measurement error
or  contamination by other macro variables unconnected to asset prices.
Denote the variance matrix of $\epsilon _{t}$ by $\Sigma _{\epsilon }$. If $%
\delta $ is a full-rank square matrix, then one can show that proxies $F_{t}$
can price assets as well as the true risk factors $G_{t}$. Indeed,
\begin{shrinkeq}{-2ex}
\begin{equation*}
\beta _{i}=\mathrm{var}(F_{t})^{-1}\mathrm{cov}(F_{t},r_{it})=(\delta \Sigma _{G}\delta
^{\prime }+\Sigma _{\epsilon })^{-1}\delta \Sigma _{G}\beta _{i}^{G}=A\beta
_{i}^{G},
\end{equation*}
\end{shrinkeq}%
where $A=(\delta \Sigma _{G}\delta ^{\prime }+\Sigma _{\epsilon
})^{-1}\delta \Sigma _{G}$ is a full-rank square matrix. Thus,
\begin{shrinkeq}{-2ex}
\begin{equation*}
Er_{it}=EG_{t}\beta _{i}^{G}=\lambda \beta _{i},
\end{equation*}
\end{shrinkeq}%
where $\lambda =A^{-1}EG_{t}$. So we see that if the econometrician is
estimating a linear factor-pricing model using factors $F_{t},$ she
has a correctly-specified model; however, this model (unlike the model with
the observed factors $G_{t}$) has a factor structure in its error terms.
Indeed, using some simple algebra one can show that equation (\ref{eq:
initial FMB model}) holds with
\begin{shrinkeq}{-2ex}
\begin{equation*}
\varepsilon _{it}=(\Sigma _{\epsilon }\delta ^{-1}\Sigma
_{G}^{-1}(G_{t}-EG_{t})-\epsilon _{t})^{\prime }\beta _{i}+\varepsilon
_{it}^{G}=v_{t}^{\prime }\mu _{i}+\varepsilon _{it}^{G}.
\end{equation*}
\end{shrinkeq}%
What is interesting here is that while the factors $v_{t}$ (and the errors $%
\varepsilon _{it}$ themselves) are uncorrelated with the observed factors $%
F_{t}$, the loadings on the error factors, $\mu _{i}$'s, and the original
loadings, $\beta _{i}$'s, are closely related (in this particular case $\mu
_{i}=\beta _{i}$). We will make use of this observation in our discussion of the
validity of the two-pass procedure. $\square $

\medskip

\textbf{Example 2.}
Consider a situation when one of the risk factors driving asset returns is
fully arbitraged and thus carries a zero risk premium. If an
econometrician does not observe this factor but does have observations on
all other relevant risk factors, then her linear factor-pricing model that
omits the arbitraged factor may still be correctly specified, while the
arbitraged factor is moved to the error term, resulting in a missing-factor
structure in the errors. $\square $

\medskip

Kleibergen and Zhan (2015) provide numerous pieces of empirical evidence
that residuals from many well-known estimated linear factor-pricing models
have non-trivial factor structures. For example, they point out that the
first three principle components of the residuals from different
pricing-model specifications used in the seminal paper by Lettau and
Ludvigson (2001) explain from 82\% to 95\% of all residual
variation. They also show that the
largest eigenvalue of the covariance matrix of residuals in all these
examples is very large and strongly separated from other eigenvalues that
are bunched together. Combining these results with the theoretical results on the limiting
distribution of  eigenvalues from Onatski (2012), one would suspect there
is at least one strong factor present in the residuals. At least five
other prominent factor-pricing studies cited in Kleibergen and Zhan (2015)
demonstrate similar evidence of  strong factor structures left in the residuals.

\medskip

\textbf{Relation between factor structure and correct specification.}
One may wonder whether the fact that the errors $\varepsilon_{it}$ in
model (\ref{eq: initial FMB model}) have a factor structure implies that the
pricing model is misspecified. The answer is ``no''; the linear
factor pricing model describes the expectations of excess returns, while the
factor structure in the errors is related to their covariances or co-movements.
It is easy to see that if the risk exposure and risk
premia on the variables $F_t$ price the assets, then the variables $F_t$
co-move the assets' returns and produce  factor-structure dependence in the
returns. However, not all co-movements of returns must carry  non-zero risk
premia; those co-movements can be placed in the error term without causing
misspecification of the pricing model.

The  correct specification of a pricing model
requires keeping  those pricing factors $F_{t,2}$ that carry
small coefficients of exposure $\beta_{2,t}$ and produce only a weak factor
structure in returns in the model. Dropping such observed factors from the
specification  leads to
asymptotically misleading inferences for both the two-pass procedure and our
proposed procedure.

\textbf{Remark.} The literature on factor pricing distinguishes cases
of tradable and non-tradable factors.  If a specific factor $F_{t}$ is a tradable portfolio and  is
supposed to be priced by the same pricing model, then $\lambda =EF_{t}$, and one can get an estimate of risk premia as the sample average of the excess returns. However, even in this case there is a value from having an alternative estimator based on the pricing equation (1). Lewellen, Nagel and Shanken (2010) showed that it is relatively easy to price the market with high cross-sectional $R^2$ by any set of portfolios as long as their number is larger than the number of the main principle components, but only if one does not enforce that the risk premia be equal the average returns. The cross-sectional $R^2$ of the pricing model is much smaller if one enforces such a restriction. Thus, having an estimator of the risk premia coming from the two-step pricing logic and comparing it to the average excess return for tradable factors is a valuable test of the pricing model.

\section{Setup and assumptions}

\label{section- assumptions}

\subsection{Model}

We consider the problem of estimation and inference on the risk premia $%
\lambda $ based on observations of returns $\{r_{it},$ $i=1,...,N,$ $%
t=1,...,T\}$ and factors $\{F_{t},$ $t=1,...,T\}$ obeying a correctly-specified factor-pricing model:
\begin{shrinkeq}{-1ex}
\begin{equation}
r_{it}=\lambda ^{\prime }\beta _{i}+(F_{t}-EF_{t})^{\prime }\beta
_{i}+v_{t}^{\prime }\mu _{i}+e_{it},  \label{eq: main formulation of dgp}
\end{equation}
\end{shrinkeq}%
where the random unobserved factor $%
v_{t}$ has zero mean and is uncorrelated with $F_{t}$. The idiosyncratic
error terms $e_{it}$ also have zero mean and are uncorrelated with $F_{t}$ and  $v_{t}$. Denote by $\mathcal{F} $
 the sigma-algebra generated by the random variables $\left(
F_{1},...,F_{T}\right) $ and $\left(
v_{1},...,v_{T}\right) $. Let $\gamma _{i}^{\prime }=\left( \beta
_{1i}^{\prime },\sqrt{T}\beta _{2i}^{\prime },\mu _{i}^{\prime }\right) $
and $\Gamma _{N}^{\prime }=\left( \gamma _{1},...,\gamma _{N}\right) $ be the $k\times N$ matrix, where $k=k_F+k_v$. Technically, $\gamma_{i,N,T}$ is more accurate indexing, as parameters $\gamma_i$ may  change with the sample size as do all other features of the data generating process, but we will drop $N,T$ to reduce clutter.

\medskip

\noindent\textbf{Assumption FACTORS.}
The $k_{F}\times 1$ vector of observed factors $F_{t}$ is stationary with finite
fourth moments and a full-rank covariance matrix $\Sigma_F$. The $k_v\times 1$
vector of unobserved factors $v_t$ is such that the following asymptotic statements hold jointly:
\begin{shrinkeq}{-1ex}
\begin{align*}
\left(
  \begin{array}{c}
    \frac{1}{\sqrt{T}}\sum_{t=1}^T(F_t-EF_t) \\
    \eta _{T}=\frac{1}{\sqrt{T}}\sum_{t=1}^{T}\Sigma _{F}^{-1}%
\widetilde{F}_{t}v_{t}^{\prime } \\
    \eta _{v,T}=\frac{1}{\sqrt{T}}\sum_{t=1}^{T}v_{t} \\
  \end{array}
\right)\Rightarrow
\left(
  \begin{array}{c}
    N(0,\Omega_F) \\
    \eta \\
    \eta _{v} \\
  \end{array}
\right),
\end{align*}
\end{shrinkeq}%
where $\mathrm{vec}(\eta )\sim N
\left( 0_{k_{F}k_{v },1},\Omega _{vF}\right) $, $\eta _{v}\sim N(0_{k_{v},1},I_{k_{v}})$ and $\widetilde{F}_t=F_t-\frac{1}{T}\sum_{s=1}^TF_s$.

\medskip

\noindent\textbf{Assumption LOADINGS.}
As both $N$ and $T$ increase to infinity, we have $N^{-1}\Gamma _{N}^{\prime
}\Gamma _{N}\rightarrow \Gamma $, where $\Gamma $ is a
positive definite $k\times k$ matrix. Also assume that $\max_{N,T}\frac{1}{N}\sum_{i=1}^N\Vert \gamma_i\Vert^4<\infty.$

\medskip

\noindent\textbf{Assumptions ERRORS.}
\begin{itemize}
\item[(i)] Conditional on $\mathcal{F}$, the random vectors $%
e_{t}=(e_{1t},...,e_{Nt})^{\prime }$ are serially independent, and $E(e_{t}|%
\mathcal{F})=0$ for all $t$.

\item[(ii)] Let $\rho (t,s)=\frac{1}{\sqrt{N}}\sum_{i=1}^{N}e_{it}e_{is}.$
Assume $\sup_{t}\sup_{s\neq t}E\left[ (1+\Vert F_{t}\Vert ^{4})(\rho
(s,t)^{2}+1)\right] <C.$

\item[(iii)] Let $S_{t}=\frac{1}{N}\sum_{i=1}^{N}e_{it}^{2}.$ Assume $\frac{%
\sqrt{N} }{T}\sum_{t=1}^{T}\widetilde{F}_{t}S_{t}=o_p(1)$ and $\frac{1}{T}%
\sum_{t=1}^{T}\widetilde{F}_{t}\widetilde{F}_{t}^{\prime }S_{t}\rightarrow
^{p}\Sigma _{SF^{2}}$.

\item[(iv)] Let $W_{t}=\frac{1}{\sqrt{N}}\sum_{i=1}^{N}\gamma_ie_{it}$. Assume   $E%
\left[ (1+\Vert F_{t}\Vert ^{2 })\Vert W_{t}\Vert ^{2}\right]
<\infty .$
\end{itemize}

\subsection{Discussion of Assumptions}

\label{subsection-discussion of assumptions}

This paper adopts an asymptotic setting when both $N$ and $T$ increase to infinity without any restriction on  relative size. Obviously, in an application we have fixed values of $N$ and $T$, and the asymptotic framework only provides an approximation to the distribution of different statistics. A natural question is when the approximations are good enough and what specific features of the application the different asymptotic rates are trying to replicate. We employ Law of Large Numbers and Central Limit Theorems for averages where summation happens over both time series and cross-sectional indexes. There are trade-offs between how large $N$ and $T$ are needed for a good approximation and the dependence structure and/or moment restrictions. Assumption ERRORS are quite agnostic about cross-sectional dependence structure but if one imposes stricter restrictions on idiosyncratic errors, then one can get good approximations even when only one index is large. Under stricter dependence assumptions, one may prove  analogues of all our statements (for example, consistency of our proposed estimator) assuming that $N\to\infty$ while $T$ is fixed, if the object of interest is ex-post risk premia, $\widetilde\lambda$, as defined in Shanken (1992).

\textbf{Assumption FACTORS.} Since  $v_t$ is a structural part of the error term  we already assume it is uncorrelated with $\widetilde{F}_t$. One can come up with a variety of  assumptions
on decaying dependence and moment conditions that would guarantee some version of the Central Limit Theorem as stated in Assumption FACTORS. The restriction that the
asymptotic covariance matrix of $\eta_v$ be the identity matrix is without a loss of generality and is just a normalization, as
neither $v_t$ nor loadings $\mu_i$ are observed.

\textbf{Assumption LOADINGS.}
In this paper we treat the loadings $\beta _{i}$ and $\mu _{i}$ as unknown
non-random vectors, the true values of which may change with the sample
sizes $N$ and $T$. Assumption LOADINGS
characterizes the size of the loadings as the sample size increases. Notice
that the loadings on the factors $F_{t,1}$ and $v_{t}$ are treated
differently than the loadings on $F_{t,2}$. Following Onatski (2012), we will
refer to the former as \textquotedblleft strong factors\textquotedblright\
and the latter as \textquotedblleft weak factors.\textquotedblright\ The
cross-sectional average of squared loadings is closely connected to the
explanatory power the factors exhibit in cross-sectional variation. The
assumptions we make on the loadings $\beta _{i,1}$ and $\mu _{i}$ guarantee
that the explanatory power of the factors $F_{t,1}$ and $v_{t}$ dominates
that of the idiosyncratic error terms. The average squared loading on the
factor $F_{2,t}$, however, converges to zero at the rate $1/T$; if $N$ and $%
T $ increase proportionally, this will lead to factor $F_{2,t}$ having
explanatory power comparable to that of the idiosyncratic errors. One
characteristic of a weak factor is the following: if it is not
observed we can  neither  consistently estimate it via the method of
principle components  nor  consistently detect it.

The loadings $\beta _{i,2}$ are asymptotically of the same order of
magnitude as $\beta _{i,1}$ divided by $\sqrt{T}$. Assumption LOADINGS makes that
the standard deviation of the first-step estimate $\widehat{\beta }_{i,2}$  be
of the same order of magnitude as $\beta _{i,2}$ itself. As we show,
this is enough to make the two-pass estimator of the risk premia $\lambda
_{2}$ on the weak factor $F_{t,2}$ inconsistent and to invalidate the
classical confidence interval for the risk premia $\lambda _{1}$ on the
strong factor $F_{t,1}$. The modeling assumption that makes $\beta _{i,2}$
drift to zero at the $\sqrt{T}$ rate is similar to assumptions
Kleibergen (2009) makes.

It is also important that the assumption on loadings $\mu _{i}$ are such that
the unobserved factor $v_{t}$ in the error terms is strong. This is
consistent with the empirical observations in Kleibergen and Zhan (2015). This also guarantees that the presence of the factor structure
plays an important role in the asymptotics of two-pass estimation.

\textbf{Assumption ERRORS.}
Assumptions ERRORS are high-level assumptions, whose main goal is to allow
for very flexible weak cross-sectional dependence among the idiosyncratic
errors, as well as flexible conditional heteroscedasticity and dependence in
higher-order moments of errors and factors. The random variables $\rho (s,t)$
stand for a (normalized) empirical analog of the error autocorrelation
coefficient, $S_{t}$ is an empirical variance, and $W_{t}$ is a (normalized)
weighted average error. These variables are normalized so that they are
stochastically bounded when the errors are cross-sectionally i.i.d.

Serial independence of errors as stated in Assumption ERRORS(i) is
consistent with the efficient market hypothesis and the unpredictability of
asset returns, and is generally consistent with empirical evidence and
tradition in the literature. This assumption may be weakened, though we do
not pursue this in the current paper.

In order to clarify the content of Assumptions ERRORS and to show that our assumptions are more flexible than those typically made in the literature, first, we provide below a set of more
restrictive primitive assumptions that are common in the literature and that
guarantee the validity of  Assumptions ERRORS. Secondly, we also provide
an empirically relevant example  not covered by the primitive
assumptions  but which satisfies our more general Assumptions ERRORS.

\medskip

\noindent\textbf{Assumptions ERRORS$^*$}
\begin{itemize}
\item[(i)] The factors $\{F_{t},$ $t=1,...,T\}$ are independent from the errors $%
\{e_{it},$ $i=1,...,N,$ $t=1,...,T\}$; the error terms $%
e_{t}=(e_{1t},...,e_{Nt})^{\prime }$ are serially independent and
identically distributed for different $t$ with $Ee_{it}=0 $ and $%
\sup_{i,t}Ee_{it}^{4}<\infty $.

\item[(ii)] Let $\mathcal{E}_{N,T}=E\left[ e_{t}e_{t}^{\prime }\right] $ be
the $N\times N$ cross-sectional covariance matrix. For some positive
constants $a,$ $c$ and $C,$ we have $\lim_{N,T}\frac{1}{N}\mathrm{tr}(\mathcal{E}_{N,T})=a$ and
\begin{shrinkeq}{-2ex}
\begin{equation*}
c< \liminf_{N,T\rightarrow \infty }\min eval\left( \mathcal{E}_{N,T}\right)
<\limsup_{N,T\rightarrow \infty }\max eval\left( \mathcal{E}_{N,T}\right) <C.
\end{equation*}
\end{shrinkeq}

\vspace{-3mm}
\item[(iii)]  $E\left|\frac{1}{\sqrt{N}}%
\sum_{i=1}^{N}(e_{it}^2-Ee_{it}^2)\right|^2<C$.
\end{itemize}

\begin{lemma}
\label{lemma- assumptions ok} Assumptions LOADINGS and  ERRORS$%
^{\ast }$ imply Assumptions ERRORS.
\end{lemma}

The primitive Assumptions ERRORS$^{\ast }$ are very close to those standard
in the literature. Numerous papers that establish inferences in factor
models commonly assume that the set of factors $\{F_{t},$ $t=1,...,T\}$
is independent from the set $\{e_{it},$ $i=1,...,N,$ $t=1,...,T\}$, though
cross-sectional dependence of errors  is allowed; see, for example, Assumption D in Bai
and Ng (2006). Many papers allow for both time-series and cross-sectional
error dependence. We exclude time-series dependence, which is justified by
the efficient-market hypothesis in our application. Assumption ERRORS$^{\ast
}$(ii) is intended to impose only weak dependence cross-sectionally as
expressed by the covariance matrix; similar assumptions appear in Onatski
(2012) and in Bai and Ng (2006).

Our high-level Assumptions ERRORS are more general than the more
standard primitive Assumptions ERRORS$^{\ast }$. In particular, our assumptions allow
for very flexible conditional heteroscedasticity in the error terms and
time-varying cross-sectional dependence, which seems  relevant when we
consider observed factors that characterize market conditions like the
momentum factor. Consider the following example.

\medskip

\noindent\textbf{Example 3.}
Assume that  errors $e_{it}$ have the following weak latent factor
structure:
\begin{shrinkeq}{-2ex}
\begin{equation*}
e_{it}=\pi _{i}^{\prime }w_{t}+\eta _{it},
\end{equation*}
\end{shrinkeq}%
where $(w_{t},F_{t})$ is stationary, $%
w_{t}$ is  $k_{w}\times 1$, serially independent conditional on $\mathcal{F}
$, with $E(w_{t}|\mathcal{F})=0$ and $E(w_{t}w_{t}^{\prime
})=I_{k_{w}}$ (normalization). Assume $E\left[ (\Vert F_{t}\Vert ^{4}+1)(\Vert w_{t}\Vert
^{4}+1)\right] <\infty .$ We assume that the loadings satisfy the condition $%
\sum_{i=1}^{N}\pi _{i}\pi _{i}^{\prime }\rightarrow \Gamma _{\pi }$ (the
factors $w_{t}$ are weak), and $N^{-1/2}\sum_{i=1}^N\pi _{i}\gamma _{i}^{\prime
}\rightarrow \Gamma _{\pi \gamma }$. Assume that the random variables $\eta _{it}$ are
independent both cross-sectionally and across time,  are independent from
$w_{t}$ and $F_{t},$ and have mean zero and finite fourth moments and variances $%
\sigma _{i}^{2}$ that are bounded above and are such that $N^{-1}\sum_{i=1}^N\sigma
_{i}^{2}\rightarrow \sigma ^{2}$. As proven in the Supplementary Appendix, this example
satisfies Assumptions ERRORS.

An interesting feature of this example is that it allows the errors to be
weakly cross-sectionally dependent to the extent that they may possess a weak
factor structure. Moreover, this factor structure may be closely related to
the observed factors $F_{t}$, which causes the cross-sectional dependence
among the errors $e_{it}$ to change with the observed factors $F_{t}$ and
allows a very flexible form of conditional heteroskedasicity. Indeed, the
conditional cross-sectional covariance is
\begin{shrinkeq}{-2ex}
\begin{equation*}
E(e_{it}e_{jt}|\mathcal{F})=\pi _{i}^{\prime }E(w_{t}w_{t}^{\prime }|%
\mathcal{F})\pi _{j}+\mathbb{I}_{\{i=j\}}\sigma _{i}^{2}.
\end{equation*}
\end{shrinkeq}%
Since we do not restrict $E(w_{t}w_{t}^{\prime }|\mathcal{F})$ beyond the
proper moment conditions, the strength of any cross-sectional dependence as well
as error variances may change stochastically  depending on the realizations
of the observed factors. This flexibility is extremely relevant for
observed factors such as the momentum. For example, one may consider $%
w_{t}=\varsigma _{t}g(F_{t},F_{t-1},...),$ where $\varsigma _{t}\sim N(0,1)$
is independent from all other variables; then for a proper choice of the
function $g(\cdot )$ one may observe higher volatility and cross-sectional
dependence of the idiosyncratic error for higher values of the observed
factor $F_{t}$.

\section{Asymptotic properties of the two-pass procedure}
\label{section- FMB fail}

In this section we derive a result concerning the asymptotic properties of the
classical two-pass procedure in a correctly-specified factor pricing model that may or may not include
weak observed factors and may or may not have missing factors in the
errors. Let us introduce the following notation: $\widetilde{F}_t=F_t-\frac{1}{T}\sum_{s=1}^TF_s$ and
\begin{shrinkeq}{-1ex}
\begin{equation*}
\widetilde{\lambda }=\lambda +\frac{1}{T}\sum_{t=1}^{T}F_{t}-EF_{t},~~~u_{i}=%
\frac{1}{T}\sum_{t=1}^{T}\Sigma _{F}^{-1}\widetilde{F}_{t}e_{it}.
\end{equation*}
\end{shrinkeq}%
Random quantity $\widetilde\lambda$ is also known as ex-post risk premia, it was introduced in Shanken (1992) and is the object of interest in Raponi, Robotti and Zaffaroni (2017) and Kim and Skoulakis (2018).

Now let us introduce two asymptotically important terms, the meaning and the
names of which will be explained in the discussion following Theorem \ref%
{theorem: FMB fails}. The first term we refer to as \textquotedblleft attenuation
bias\textquotedblright \ is
\begin{equation*}
B^A=-\left( \sum_{i=1}^{N}\widehat{\beta }_{i}\widehat{\beta }_{i}^{\prime
}\right) ^{-1}\sum_{i=1}^{N}u_{i}u_{i}^{\prime }\widetilde{\lambda },
\end{equation*}
while the second term we will call the \textquotedblleft omitted variable
bias\textquotedblright \ is
\begin{equation*}
B^{OV}=\left( \sum_{i=1}^{N}\widehat{\beta }_{i}\widehat{\beta }_{i}^{\prime
}\right) ^{-1}\sum_{i=1}^{N}\widehat{\beta }_{i}\frac{\mu _{i}^{\prime }}{%
\sqrt{T}}(\eta _{v,T}-\eta _{T}^{\prime }\widetilde{\lambda }).
\end{equation*}
These terms are not biases in an exact sense as they are random, but rather
they are sample analogues of the expressions that are classically known as
attenuation and omitted variable biases. Notice that both quantities are
infeasible as they depend
on unobserved errors $e_{it}$, unobserved factors $v_{t}$ and unknown
parameters $\lambda $ and $\mu _{i}$. Both terms are $k_{F}\times 1$
vectors. Let $B^A_{1}$ and $B^{OV}_{1}$ denote $k_{1}\times 1$ sub-vectors
containing the first $k_{1}$ components, while $B^A_{2}$ and $B^{OV}_{2}$ are
$k_{2}\times 1$ sub-vectors of the last $k_{2}$ components of $B^A$ and $B^{OV}$,
correspondingly. We also adopt the following notation: $\Gamma _{\beta _{2}\mu }$ is the $%
k_{2}\times k_{\mu }$ sub-block of matrix $\Gamma $ (defined in Assumption LOADINGS) corresponding
to the limit of $N^{-1}\sum_{i=1}^{N}\sqrt{T}\beta _{i,2}\mu _{i}$. Other
sub-matrices are denoted similarly.

\begin{theorem}
\label{theorem: FMB fails}Assume that the samples $\{r_{it},$ $i=1,...,N,$ $%
t=1,...,T\}$ and $\{F_{t},$ $t=1,...,T\}$ come from a data-generating
process that satisfies  factor-pricing model (\ref{eq: main formulation
of dgp}) and assumptions FACTORS, LOADINGS and
ERRORS. Let $\widehat{\lambda }_{TP}$ denote the estimate obtained via the
conventional two-pass procedure. Let both $N$ and $T$ increase to infinity
without restrictions on relative rates. Then the following asymptotic
statements hold simultaneously:
\begin{align*}
\left(
\begin{array}{c}
\sqrt{T}B^{OV}_{1} \\
B^{OV}_{2}%
\end{array}%
\right) & \Rightarrow \left( (I_{k_{\beta }};\widetilde{\eta })
\Gamma (I_{k_{\beta }};\widetilde{\eta })^{\prime }+\mathcal{I}%
_{k_{2}}\Sigma _{u}\mathcal{I}_{k_{2}}\right) ^{-1}\left( \Gamma
_{\beta \mu }+\widetilde{\eta }\Gamma_{\mu \mu }\right) (\eta
_{v}-\eta ^{\prime }\lambda ), \\
\left(
\begin{array}{c}
\sqrt{T}B^A_{1} \\
B^A_{2}%
\end{array}%
\right) & \Rightarrow -\left( (I_{k_{\beta }};\widetilde{\eta })
\Gamma (I_{k_{\beta }};\widetilde{\eta })^{\prime }+\mathcal{I}%
_{k_{2}}\Sigma _{u}\mathcal{I}_{k_{2}}\right) ^{-1}\mathcal{I}_{k_{2}}\Sigma
_{u}\lambda , \\
\sqrt{T}(\widetilde{\lambda }-\lambda )& \Rightarrow N(0,\Omega _{F}),
\end{align*}%
and
\begin{equation*}
\left(
\begin{array}{c}
\sqrt{NT}(\widehat{\lambda }_{TP,1}-\widetilde{\lambda }_{1}-B^A_{1}-B^{OV}_{1})
\\
\sqrt{N}(\widehat{\lambda }_{TP,2}-\widetilde{\lambda }_{2}-B^A_{2}-B^{OV}_{2})%
\end{array}%
\right) =O_{p}(1),
\end{equation*}%
where $\Sigma _{u}=\Sigma_F^{-1}\Sigma_{SF^2}\Sigma_F^{-1}$,  $\mathcal{I}_{k_{2}}=\left(
\begin{array}{cc}
0_{k_{1},k_{1}} & 0_{k_{1},k_{2}} \\
0_{k_{2},k_{1}} & I_{k_{2}}%
\end{array}%
\right) $ is a $k_{F}\times k_{F}$ matrix, and $\widetilde{\eta }=\mathcal{I}%
_{k_{2}}\eta $ is a $k_{F}\times k_{v}$ random matrix (with $\eta $
as in Assumption FACTORS).
\end{theorem}

Theorem \ref{theorem: FMB fails} states the rates of convergence for
different parts of the two-pass estimator. Notice that the theorem does not
impose a relative rate of increase between $N$ and $T$ as long as both
increase to infinity simultaneously. One observation is that the two-pass
procedure cannot estimate $\lambda $ at a rate faster than $\sqrt{T}$
despite the fact that the dataset has $NT$ observations of portfolio excess
returns, and one could expect the $\sqrt{NT}$ rate. This arises from the fact
that the correct specification  (\ref{eq: initial FMB model}) if averaged across time,
gives
\begin{shrinkeq}{-2ex}
\begin{equation}\label{eq: ideal cross sectional}
\overline{r}_{i}=\widetilde{\lambda }\beta _{i}+\overline{\varepsilon }_{i}.
\end{equation}
\end{shrinkeq}%
Thus, even if $\beta _{i}$ were known, the `true' coefficient $\widetilde{%
\lambda }$ in the only ideal regression we have (that is, regression of
average return on $\beta _{i}$) differs from the parameter $\lambda $ we
want to estimate, by the term $\frac{1}{T}\sum_{t=1}^{T}F_{t}-EF_{t}$,
which, if multiplied by $\sqrt{T}$, is asymptotically zero mean gaussian
with variance $\Omega _{F}$. Notice that if all observed factors $F_{t}$ are
excess returns themselves and are assumed to be priced by the same pricing
model, then the asset pricing theory provides an alternative way of
estimating risk premia. Namely, in such a case $\lambda =EF_{t},$ and the
alternative estimate is $\widehat{\lambda }=\frac{1}{T}\sum_{t=1}^{T}F_{t}=%
\widetilde{\lambda }$. However, this estimate is not valid if factors
themselves are not excess returns or are not priced by the same model.

Notice also that if the limits of the normalized $B^{OV}$ and $B^A$ are
non-zero, then these terms (together with $\widetilde{\lambda }_{1}$)
asymptotically dominate estimation. Below we consider three cases
covered by Theorem \ref{theorem: FMB fails}. The first one is the case with
no weak observed factors ($k_{2}=0$). In this case the theorem delivers the
validity of the two-pass procedure, namely, that the two-pass estimator is consistent and
asymptotically mean-zero gaussian. For the other two (more empirically
relevant) cases -- one with weak observed factors but no missing factors,
the other with weak observed factors and missing strong factors -- the
two-pass procedure fails.

\subsection{No weak observed factors}

\begin{corollary}
\label{cor: FMB works} Assume that the samples $\{r_{it},$ $i=1,...,N,$ $
t=1,...,T\}$ and $\{F_{t},$ $t=1,...,T\}$ come from a data-generating
process that satisfies  factor-pricing model (\ref{eq: main formulation of
dgp}) and assumptions FACTORS, LOADINGS and ERRORS
with $k_{2}=0$ (no weak observed factors). Then,
\vspace{-3mm}
\begin{equation*}
\sqrt{T}(\widehat{\lambda }_{TP}-\lambda )\Rightarrow \Gamma_{\beta \beta }^{-1}\Gamma_{\beta \mu }(\eta _{v}-\eta ^{\prime
}\lambda )+\lim \sqrt{T}(\widetilde{\lambda }-\lambda ),
\end{equation*}
where the limit on the right hand side is asymptotically gaussian with mean
zero. If in addition there are no strong missing factors in the errors
(that is, $\mu _{i}=0$), then
\begin{shrinkeq}{-1ex}\begin{equation*}
\sqrt{T}(\widehat{\lambda }_{TP}-\lambda )=\sqrt{T}(\widetilde{\lambda }%
-\lambda )+o_{p}(1)\Rightarrow N(0,\Omega _{F}).
\end{equation*}\end{shrinkeq}
\end{corollary}
This is a positive statement about the two-pass procedure, which claims that if
all observed factors are strong, then the two-pass procedure is $\sqrt{T}$%
-consistent and provides an asymptotically mean-zero gaussian estimate for the risk premia when
both $N,T\rightarrow \infty $. If the error terms have a strong factor
structure it does not lead to a bias, but may increase the asymptotic
variance. If no strong missing factor structure is present in the error terms, then
the two-pass procedure is asymptotically equivalent to the infeasible
estimate $\widetilde{\lambda }$ and will have asymptotic variance $\Omega _{F}$.

\subsection{Weak observed factors but no strong missing factors}

\begin{corollary}
\label{theorem: FMB fails no missing} Assume that the samples $\{r_{it},$ $i=1,...,N,$ $
t=1,...,T\}$ and $\{F_{t},$ $t=1,...,T\}$ come from a
data-generating process that satisfies  factor-pricing model (\ref{eq: main
formulation of dgp}) and assumptions FACTORS, LOADINGS and
ERRORS with $k_{2}\geq 1$ (there are weak observed factors) and $%
k_{v}=0$ (no missing factor structure in errors). Then the following
asymptotic statements hold jointly:
\begin{shrinkeq}{-2ex}
\begin{align*}
\sqrt{T}(\widehat{\lambda }_{TP,1}-\lambda _{1})& =\sqrt{T}(\widetilde{%
\lambda }_{1}-\lambda _{1})+\sqrt{T}B^A_{1}+o_{p}(1), \\
\widehat{\lambda }_{TP,2}-\lambda _{2}& =B^A_{2}+o_{p}(1),
\end{align*}
\end{shrinkeq}%
where
\begin{shrinkeq}{-1ex}
\begin{equation}
\left(
\begin{array}{c}
\sqrt{T}B^A_{1} \\
B^A_{2}%
\end{array}%
\right) \rightarrow ^{p}-\left( \Gamma+\mathcal{I}_{k_{2}}\Sigma
_{u}\mathcal{I}_{k_{2}}\right) ^{-1}\mathcal{I}_{k_{2}}\Sigma _{u}\lambda .
\label{eq: ab in no missing case}
\end{equation}
\end{shrinkeq}%
\end{corollary}

In the case when some of the observed factors have relatively small loadings
(weak observed factors) the two-pass estimator will deviate from the classical
case even if the idiosyncratic errors are not strongly correlated. The limit
in equation (\ref{eq: ab in no missing case})  is a non-random,  non-zero vector,
and thus characterizes the asymptotic bias. The two-pass estimate $\widehat{%
\lambda }_{TP,2}$ of the risk premia on weak factors $F_{t,2}$ is
inconsistent and converges in probability to an incorrect  value. The two-pass
estimate $\widehat{\lambda }_{TP,1}$ of risk premia on strong factors $F_{t,1}$
is $\sqrt{T}$-consistent, but  has a bias of order $\frac{1}{%
\sqrt{T}}$, the same order of magnitude as the standard deviation of its
asymptotic distribution. This leads to  invalid inferences on the risk premia.

The result of Corollary \ref{theorem: FMB fails no missing} can be explained
in terms of classical error-in-variables bias. The first-pass estimate $\widehat{\beta }_{i}$ of risk
exposure coefficients $\beta _{i}$ contains estimation errors which are
stochastically of order $O_{p}(1/\sqrt{T})$ each:
\begin{shrinkeq}{-1ex}
\begin{equation*}
\widehat{\beta }_{i}=\left( \sum_{t=1}^{T}\widetilde{F}_{t}\widetilde{F}%
_{t}^{\prime }\right) ^{-1}\sum_{t=1}^{T}\widetilde{F}_{t}r_{it}=(\beta
_{i}+u_{i})(1+o_{p}(1)),
\end{equation*}\end{shrinkeq}%
where the $o_{p}(1)$ term is related to the difference between $\Sigma
_{F}=E[(F_{t}-EF_{t})(F_{t}-EF_{t})^{\prime }]$ and $T^{-1}\sum_{t}\widetilde{F}%
_{t}\widetilde{F}_{t}^{\prime }$. As a result, the second-pass regression
encounters an error-in-variables problem. In the case of exposure to a
strong observed factor, the estimation error in $\widehat{\beta }_{i,1}$ is
asymptotically negligible compared to the size of the coefficient $\beta
_{i,1}$ itself, and so this estimation error does not jeopardize
consistency. However, the estimation error in $\widehat{\beta }_{i,2}$ is
asymptotically of the same order of magnitude as the coefficient itself. The
first-pass estimation errors in $\widehat{\beta }_{i,2}$ behave like a
classical measurement error in the following sense: the imposed assumptions
guarantee that the estimation errors $u_{i,2}$ for different assets are
asymptotically uncorrelated and that they are asymptotically uncorrelated
with $\beta _{i}$
themselves in the sense that the sample correlation between $\beta _{i}$ and
$u_{i}$ is asymptotically negligible. The bias we observe in Corollary \ref%
{theorem: FMB fails no missing} is a classical attenuation bias, with $%
\mathcal{I}_{k_{2}}\Sigma _{u}\mathcal{I}_{k_{2}}$ corresponding to the
variance of the normalized measurement error $\sqrt{T}u_{i,2}$.

\textbf{Remark.} A result  similar to Corollary 2 can be derived under assumptions that $N$ is fixed while $T\to\infty$ and is due to Kleibergen (2009).

Note that if $\Gamma=\Gamma_{\beta\beta}$ is a block
diagonal matrix with $\Gamma_{\beta _{1}\beta
_{2}}=0_{k_{1},k_{2}},$ the two-pass procedure inferences about $\lambda _{1}
$ will not be disturbed; namely, $\widehat{\lambda }_{TP,1}$ will be $\sqrt{T%
}$-consistent and  will have an asymptotically
mean-zero gaussian distribution. The block-diagonality assumption, however,
is a very strong one: it requires that the values of $\beta _{i,1}$ be
unrelated to the values of $\beta _{i,2}$ for the same asset, which is both
implausible and not supported in applications. For example, the sample
correlation coefficient between portfolios' betas that correspond to the
market portfolio and the betas that correspond to the SMB (HML) portfolio in the
Fama--French dataset are equal to 0.73 (0.47).

\subsection{Weak observed factors and strong missing factors}

\begin{corollary}
\label{cor: general case} Assume that the sample $\{r_{it},$ $i=1,...,N,$ $%
t=1,...,T\}$ and $\{F_{t},$ $t=1,...,T\}$ comes from a data-generating
process that satisfies  factor-pricing model (\ref{eq: main formulation of
dgp}) and assumptions FACTORS, LOADINGS and ERRORS
with $k_{2}\geq 1$ (there are weak observed factors) and $k_{v}\geq 1$
(there is a missing factor structure in errors). Then the following
asymptotic statements hold simultaneously:
\begin{shrinkeq}{-2ex}
\begin{align*}
\sqrt{T}(\widehat{\lambda }_{TP,1}-\lambda _{1})& =\sqrt{T}(\widetilde{%
\lambda }_{1}-\lambda _{1})+\sqrt{T}B^A_{1}+\sqrt{T}B^{OV}_{1}+o_{p}(1), \\
\widehat{\lambda }_{TP,2}-\lambda _{2}& =B^A_{2}+B^{OV}_{2}+o_{p}(1),
\end{align*}
\end{shrinkeq}%
where
\begin{shrinkeq}{-2ex}
\begin{align*}
\left(
\begin{array}{c}
\sqrt{T}B^{OV}_{1} \\
B^{OV}_{2}%
\end{array}%
\right) & \Rightarrow \left( (I_{k_{\beta }};\widetilde{\eta })
\Gamma(I_{k_{\beta }};\widetilde{\eta })^{\prime }+\mathcal{I}%
_{k_{2}}\Sigma _{u}\mathcal{I}_{k_{2}}\right) ^{-1}\left( \Gamma_{\beta \mu }+\widetilde{\eta } \Gamma_{\mu \mu }\right) (\eta
_{v}-\eta ^{\prime }\lambda ), \\
\left(
\begin{array}{c}
\sqrt{T}B^A_{1} \\
B^A_{2}%
\end{array}%
\right) & \Rightarrow -\left( (I_{k_{\beta }};\widetilde{\eta })
\Gamma(I_{k_{\beta }};\widetilde{\eta })^{\prime }+\mathcal{I}%
_{k_{2}}\Sigma _{u}\mathcal{I}_{k_{2}}\right) ^{-1}\mathcal{I}_{k_{2}}\Sigma
_{u}\lambda .
\end{align*}
\end{shrinkeq}%
The distributions on the right hand side  are non-gaussian and are
not centered at zero.
\end{corollary}

This result covers a more general case which, as we argued before, is
empirically quite relevant. Here some  observed pricing factors may
have relatively small loadings (weak factors), while errors are highly
cross-sectionally correlated to the extent that they have strong missing
factor structures. The two-pass estimate $\widehat{\lambda }_{TP,2}$ of the
risk premia on weak factors $F_{t,2}$ is inconsistent and, asymptotically,
has a poorly-centered, non-standard distribution. The two-pass estimate $%
\widehat{\lambda }_{TP,1}$ of risk premia on strong factors $F_{t,1}$ is $%
\sqrt{T}$-consistent, but this estimate has a bias of order $\frac{1}{\sqrt{T}%
}$ and an asymptotically non-standard distribution. This makes standard
inferences (based on the usual $t$-statistics) invalid.

In the presence of a strong factor structure in the errors, first-pass estimates
have the following form:
\begin{shrinkeq}{-1ex}
\begin{equation}
\widehat{\beta }_{i}=\left( \sum_{t=1}^{T}\widetilde{F}_{t}\widetilde{F}%
_{t}^{\prime }\right) ^{-1}\sum_{t=1}^{T}\widetilde{F}_{t}r_{it}=\left(
\beta _{i}+\frac{\eta _{T}\mu _{i}}{\sqrt{T}}+u_{i}\right) \left(
1+o_{p}(1)\right) ,  \label{eq: first stage estimate}
\end{equation}
\end{shrinkeq}%
where $\eta _{T}=\frac{1}{\sqrt{T}}\sum_{t=1}^{T}\Sigma _{F}^{-1}\widetilde{F}%
_{t}v_{t}^{\prime }\Rightarrow \eta. $ Again, for the strong observed
factors, the estimation error in $\widehat{\beta }_{i,1}$ turns out to be asymptotically
negligible when compared to the sizes of risk exposures $\beta _{i,1}$
themselves, while the estimation errors in $\widehat{\beta }_{i,2}$ -- which
are now equal to $\eta _{T}\mu _{i}/\sqrt{T}+u_{i}$ -- are of the size $O_{p}(1/%
\sqrt{T})$, which is the same order of magnitude as the $\beta _{i,2}$'s
themselves.

The estimation errors of $\widehat{\beta }_{i,2}$ distort the asymptotics
and invalidate classical inferences. However, unlike the case covered by
Corollary \ref{theorem: FMB fails no missing}, the estimation errors in this
setting do not behave like classical measurement errors in two respects. First,
the estimation errors for different assets are correlated due to the
presence of the common component $\eta _{T}$ in all of them. Second, unless $\mu
_{i}$ is cross-sectionally uncorrelated with $\beta _{i}$ (so that $
\Gamma_{\beta \mu }=0_{k_{F},k_{v}}$), the estimation error will be
correlated with its own regressor $\beta _{i}$.

There is an additional issue  classically known as omitted variable bias. Let us look at the
second pass (normalized) `ideal' regression, which we can obtain by
time-averaging equation (\ref{eq: main formulation of dgp}):
\begin{equation}
\sqrt{T}\overline{r}_{i}=\sqrt{T}\widetilde{\lambda }^{\prime }\beta
_{i}+\eta _{v,T}^{\prime }\mu _{i}+\sqrt{T}\overline{e}_{i},
\label{eq:second step initial}
\end{equation}
where $\eta _{v,T}=\frac{1}{\sqrt{T}}\sum_{t=1}^{T}v_{t}\Rightarrow \eta
_{v}\sim N(0_{k_{v},1},I_{k_{v}})$. Here we introduced normalization $\sqrt{T%
}$ to make regression (\ref{eq:second step initial}) more compatible with the
classical OLS setup. The regression error terms $\sqrt{T}\overline{e}_{i}$
all have orders of magnitude of $O_{p}(1)$, zero means and finite variances.
Even though in finite samples $\sqrt{T}\overline{e}_{i}$ may be weakly
cross-sectionally dependent, Assumption ERRORS guarantees that they are
asymptotically uncorrelated. Imagine for a moment that we know $\beta _{i}$
and $\mu _{i}$ for all assets. Then, regression (\ref{eq:second step initial}%
) will take the form of  a classic OLS regression, with regressors $\sqrt{T}\beta _{i,2}$
and $\mu _{i}$ being of  order of magnitude  $O(1)$, in the sense
expressed in Assumption LOADINGS, that in the classical regression setting would
lead to a $\sqrt{N}$-consistent and asymptotically gaussian OLS estimator of the
coefficients on them. The regressor $\sqrt{T}%
\beta _{i,1}$ is, in contrast, of order $O(\sqrt{T})$ and carries a lot of
information which, in the classical regression setting, leads to an OLS
estimator of the coefficient $\lambda _{1}$ on this regressor that is both
super-consistent and asymptotically centered gaussian. However, because $\mu
_{i}$ is unobserved, it becomes a part of the error term in the second-pass
regression, making error terms cross-sectionally correlated; see, for
example, Andrews (2005) for a similar phenomenon. A more classical reference
for this phenomenon is that of omitted variable bias: if $\Gamma_{\beta \mu }\neq 0_{k_{F},k_{v}},$ then even if there were no first-pass
estimation error and we knew $\beta _{i}$, running an OLS in a regression of
$\sqrt{T}\overline{r}_{i}$ on $\sqrt{T}\beta _{i}$ would produce invalid
results due to the omission of $\mu _{i}$.

One question that may arise is whether or not the omitted variable bias is
large. The answer to this question is closely related to the size of the
cross-sectional correlation between $\beta _{i}$ and $\mu _{i}$ as expressed
in $\Gamma_{\beta \mu }$. Unfortunately, there is no reliable
empirical evidence on this, as $\mu _{i}$ is unobserved and $\beta _{i}$ is
poorly estimated and biased in the direction of $\mu _{i}$ (see equation (%
\ref{eq: first stage estimate})). The problem with estimation of $\mu _{i}$
is that the estimator $\widehat{\lambda }_{TP,2}$ is inconsistent, which
makes the residuals from the two-pass procedure poor indicators of the true
errors, and estimating $\mu _{i}$ via the principle components analysis on
the residuals does not produce good estimates. However, even though direct
empirical evidence on this matter is absent, we have two indirect arguments
which suggest that one should expect a high rather than low correlation
between $\beta _{i}$ and $\mu _{i}$. One argument is the empirical
observation that for many well-known factor-pricing models the estimated
betas for different factors are exceptionally highly correlated. Another
argument is related to our theoretical example 1, where the missing factor
structure originates as a result of mismeasuring the true risk factor, and
the sample correlation between $\beta _{i}$ and $\mu _{i}$ equals 1.

\section{Sample-split estimator of the risk premia}

\label{section- our method}

\subsection{Idea of the proposed solution}

\label{subsection- intuition for our solution}

\textbf{The case of no factor structure in the error terms.}
We begin by solving the easier case when no unobserved factor structure is
present in the errors, while some observed factors are weak. In such a case the failure of the two-pass
procedure can be labeled  a classical measurement error-in-variables
problem, which is often solved by finding a proper instrument. Apparently,
it is relatively easy to find a valid instrument in our setting if one is
willing to employ a sample-splitting technique.

Let us divide the set of time indexes $t=1,...,T$ into two non-intersecting
equal subsets $T_{1}$ and $T_{2}$. It is more natural to make $T_{1}$
the first half of the sample, and $T_{2}$  its second half. Let us run the
first-step regression twice -- separately on each sub-sample:
\begin{shrinkeq}{-1ex}
\begin{equation*}
\widehat{\beta }_{i}^{(j)}=\left( \sum_{t\in T_{j}}\widetilde{F}_{t}^{(j)}%
\widetilde{F}_{t}^{(j)\prime }\right) ^{-1}\sum_{t\in T_{j}}\widetilde{F}%
_{t}^{(j)}r_{it}=(\beta _{i}+u_{i}^{(j)})(1+o_{p}(1)),\mbox{  for  }j=1,2,
\end{equation*}
\end{shrinkeq}%
where $\widetilde{F}_{t}^{(j)}=F_{t}-\frac{1}{|T_{j}|}\sum_{t\in T_{j}}F_{t}$%
, $u_{i}^{(j)}=\frac{1}{|T_{j}|}\sum_{t\in T_{j}}\Sigma _{F}^{-1}\widetilde{F%
}_{t}^{(j)}e_{it}$, and the $o_{p}(1)$ term is related to the difference
between $\Sigma _{F}$ and $\frac{1%
}{|T_{j}|}\sum_{t\in T_{j}}\widetilde{F}_{t}^{(j)}\widetilde{F}_{t}^{(j)\prime }$%
.

The assumption ERRORS guarantees that the two sets of estimation
uncertainty, $\{u_{i}^{(1)},$ $i=1,...,N\}$ and $\{u_{i}^{(2)},$ $%
i=1,...,N\},$ are independent conditionally on $\mathcal{F}$. In fact, the
asymptotic independence of the two sets of errors will hold more generally
if one makes stationarity assumptions and controls the decay of time-series
dependence in errors $e_{it},$ and the sub-samples are formed to be first
and second halves of the sample.

Given the observation about independence of estimation errors obtained from
different sub-samples, one may use an estimate of $\beta _{i}$ from one
sub-sample (for example, $\widehat{\beta }_{i}^{(1)}$) as a regressor while
the other (in this example, $\widehat{\beta }_{i}^{(2)}$) as an instrument.
This would represent a valid IV regression. Indeed, the second-step
regression we run is:
\begin{shrinkeq}{-2ex}
\begin{equation*}
\overline{r}_{i}=\widetilde{\lambda }^{\prime }\widehat{\beta }_{i}^{(1)}+(%
\overline{e}_{i}-\widetilde{\lambda }^{\prime }u_{i}^{(1)}).
\end{equation*}
\end{shrinkeq}%
In this regression the regressor and the instrument are correlated since
they both contain $\beta _{i}$, hence we get the relevance condition. The
validity condition holds for two reasons: (i) the part of the second-step
regression error $u_{i}^{(1)}$ is asymptotically uncorrelated with the
instrument $\widehat{\beta }_{i}^{(2)}$; (ii) Assumption ERRORS guarantees
that $\widehat{\beta }_{i}^{(2)}$ is asymptotically uncorrelated with $%
\overline{e}_{i}.$ As we show below, this procedure restores consistency and
standard inferences on the estimates of risk premia.

Similar ideas, such as sample splitting and jackknife-type estimators, have
been previously employed in the literature on many weak instruments (e.g.,
Hansen, Hausman and Newey, 2008). In that literature the term
\textquotedblleft many instruments\textquotedblright\ is related to modeling
the number of instruments as growing to infinity proportionally to the sample size or the concentration parameter, while the
term \textquotedblleft weak\textquotedblright\ appears due to a modeling
assumption that makes the estimation error of the reduced-form coefficients
be of the same order of magnitude as the coefficients themselves (so
called local-to-zero asymptotics). This is parallel to the dimension
asymptotics for a number of portfolios and the local-to-zero asymptotics
for risk exposures of weak factors in our setup. In the many-weak-instrument
setting, the regular TSLS estimator has a significant bias,
and classical inferences are asymptotically invalid. That problem can also
be interpreted as a classical measurement-error-in-variables problem for the
second-stage regression, where the regression is run on the fitted values
from the first-stage projection of the original regressor on the
instruments. Some  proposed solutions employ the second-stage instrumental
variables regression where, for each observation, the regressor is obtained
from a first-stage regression run on a sub-sample that does not include that
observation, and the original instrument is still used as an instrument
(see Angrist, Imbens and Krueger (1999) and Dufour and Jasiak (2001)). This forces the first-stage error in the
projection  to be uncorrelated with the instrument for this specific
observation. Sample-splitting or leave-one-out type procedures restore consistency and classical inferences.

Raponi, Robotti and Zaffaroni (2017) and Kim and Skoulakis (2018) consider a similar phenomenon by assuming that $N\to\infty$ while $T$ is fixed. They show that in such a setting, the estimation error of the first stage is important and produces an inconsistent estimator of ex-post risk premia due to error-in-variable bias. These papers suggest correcting attenuation bias by directly estimating it. Those suggestions can be proved consistent in our asymptotic setting under slightly stronger Assumptions ERRORS. However, these estimation techniques would fail in the presence of missing factors in the error term.

\textbf{The case of factor structure in the error terms.}
The model with unobserved factor
structure has an additional problem -- the presence of
omitted (and unobserved) variable $\mu _{i}$ in  regression (\ref%
{eq:second step initial}). After examining  formula (\ref{eq:
first stage estimate}) for the first-pass estimate we may notice that we can
obtain a noisy proxy for $\mu _{i}$ if we take the difference between two
estimates for the same $\beta _{i}$ obtained from different sub-samples.
Indeed, consider two non-intersecting subsets of time indexes, $T_{1}$ and $%
T_{2}$, and assume  they have the same number, say $\tau ,$ of time
indexes. Then
\begin{shrinkeq}{-2ex}
\begin{equation*}
\widehat{\beta }_{i}^{(1)}-\widehat{\beta }_{i}^{(2)}=\frac{\eta _{\tau
}^{(1)}-\eta _{\tau }^{(2)}}{\sqrt{\tau }}\mu _{i}+(u_{i}^{(1)}-u_{i}^{(2)}).
\end{equation*}
\end{shrinkeq}%
Notice that both the coefficient on $\mu _{i}$ and the noise term $%
u_{i}^{(1)}-u_{i}^{(2)}$ are of the same order of magnitude $O_{p}(1/\sqrt{%
\tau })$. This means that  neither the signal dominates the noise -- and thus
we need a correction to account for the noise, -- nor the noise dominates the
signal, and thus the proxy is not useless.

Assume that $k_{v}\leq k_{F}$, which implies that we have a larger number of
proxies than needed, and we have a choice among them.  Now
we assume that we have a fixed and full-rank $k_{v}\times k_{F}$ matrix $A$,
and use $A(\widehat{\beta }_{i}^{(1)}-\widehat{\beta }_{i}^{(2)})$ as the
proxy.

The idea is to regress the average return $\overline{r}_{i}$ on $\widehat{%
\beta }_{i}^{(1)}$ and $A(\widehat{\beta }_{i}^{(1)}-\widehat{\beta }%
_{i}^{(2)})$ instead of on unobserved $\beta _{i}$ and $\mu _{i}$. This solves
the omitted-variables part of the problem, but the error-in-variables issue
still remains. That problem we solve via instrumental variables upon
additional sample splitting. The ultimate idea goes as follows: split the
sample into four equal sub-samples along the time dimension; calculate the
first-pass estimates of risk exposures for all four sub-samples; run an
instrumental variables regression using $\widehat{\beta }_{i}^{(1)}$ and $A(%
\widehat{\beta }_{i}^{(1)}-\widehat{\beta }_{i}^{(2)})$ as regressors and $%
\widehat{\beta }_{i}^{(3)}$ and $(\widehat{\beta }_{i}^{(3)}-\widehat{\beta }%
_{i}^{(4)})$ as instruments.

One can re-write the two pass-procedure as a GMM moment condition, then the two-pass will correspond to an IV estimator with many instruments, as described in  Newey and Windmeijer (2009). The problem described in Corollary 3 and the solution proposed in section 5.1 do not fall within the Newey and Windmeijer (2009) framework. The main departure is that Newey and Windmeijer (2009) consider i.i.d. sampling, while the observations in our model are so highly dependent that they produce inconsistency in the estimator.

\subsection{Algorithm for constructing four-split estimator}

\label{subsection-algorithm for estimator}

Let us divide the set of time indexes into four equal non-intersecting
subsets $T_{j},$ $j=1,...,4$.

\begin{itemize}
\item[(1)] For each asset $i$ and each subset $j$ run a time-series
regression to estimate the coefficients of risk exposure:
\begin{shrinkeq}{-2ex}
\begin{equation*}
\widehat{\beta }_{i}^{(j)}=\left( \sum_{t\in T_{j}}\widetilde{F}_{t}^{(j)}%
\widetilde{F}_{t}^{(j)\prime }\right) ^{-1}\sum_{t\in T_{j}}\widetilde{F}%
_{t}^{(j)}r_{it}.
\end{equation*}
\end{shrinkeq}

\item[(2)] Run an IV regression of $\overline{r}_{i}=\frac{1}{T}%
\sum_{t=1}^{T}r_{it}$ on regressors $x_{i}^{(1)}=\left( \widehat{\beta }%
_{i}^{(1)\prime },(\widehat{\beta }_{i}^{(1)}-\widehat{\beta }%
_{i}^{(2)})^{\prime }A_{1}^{\prime }\right) ^{\prime }$ with instruments $%
z_{i}^{(1)}=\left( \widehat{\beta }_{i}^{(3)\prime },(\widehat{\beta }%
_{i}^{(3)}-\widehat{\beta }_{i}^{(4)})^{\prime }\right) ^{\prime }$, where $A_1$ is a non-random $k_v\times k_F$ matrix of rank $k_F$. Let $\widehat{\lambda }^{(1)}$ be
the TSLS estimate of the coefficient on regressor $\widehat{\beta }%
_{i}^{(1)}. $

\item[(3)] Repeat step (2) three more times exchanging indexes 1 to 4
circularly; that is, the (2)$^{nd}$ regression is an IV regression of $\overline{r%
}_{i}$ on regressors $x_{i}^{(2)}=\left( \widehat{\beta }_{i}^{(2)\prime },(%
\widehat{\beta }_{i}^{(2)}-\widehat{\beta }_{i}^{(3)})^{\prime
}A_{2}^{\prime }\right) ^{\prime }$ with instruments $z_{i}^{(2)}=\left(
\widehat{\beta }_{i}^{(4)\prime },(\widehat{\beta }_{i}^{(4)}-\widehat{\beta
}_{i}^{(1)})^{\prime }\right) ^{\prime }$; denote the  estimate
as $\widehat{\lambda }^{(2)}$, etc.

\item[(4)] Obtain the four-split estimate as
$
\widehat{\lambda }_{4S}=\frac{1}{4}\sum_{j=1}^{4}\widehat{\lambda }^{(j)}.
$

\item[(5)] In order to compute an estimate of the covariance matrix for $%
\widehat{\lambda }_{4S},$ denote by $X^{(j)}$ the $N\times k$ matrix of
stacked regressors used in the ($j$)$^{th}$ IV regression, and by $Z^{(j)}$ the $N\times k_{z}$ matrix of instruments
from this regression (here $k_{z}=2k_{F}$  and $k=k_{F}+k_{v}$). Let $P_{Z}=Z\left( Z^{\prime }Z\right) ^{-1}Z^{\prime }$, calculate the following matrices:
\begin{shrinkeq}{-1ex}
\begin{align*}
\widehat{\Sigma }_{0}=\frac{1}{N}\sum_{i=1}^{N}\left(
\begin{array}{c}
\widetilde{z}_{i}^{(1)}\widehat{\epsilon}_{i}^{(1)} \\
... \\
\widetilde{z}_{i}^{(4)}\widehat{\epsilon}_{i}^{(4)}%
\end{array}%
\right) \left(
\begin{array}{c}
\widetilde{z}_{i}^{(1)}\widehat{\epsilon}_{i}^{(1)} \\
... \\
\widetilde{z}_{i}^{(4)}\widehat{\epsilon}_{i}^{(4)}%
\end{array}%
\right) ^{\prime },\quad G=\left(
\begin{array}{cccc}
G_{1} & 0_{k,k} & 0_{k,k} & 0_{k,k} \\
0_{k,k} & G_{2} & 0_{k,k} & 0_{k,k} \\
0_{k,k} & 0_{k,k} & G_{3} & 0_{k,k} \\
0_{k,k} & 0_{k,k} & 0_{k,k} & G_{4}%
\end{array}%
\right),
\end{align*}
\end{shrinkeq}%
where $\widehat{\epsilon}_{i}^{(j)}$ is $i^{th}$ residual from the ($j$)$^{th}$ IV regression,  $\widetilde{z}%
_{i}^{(j)}=X^{(j)\prime }Z^{(j)}(Z^{(j)\prime }Z^{(j)})^{-1}z_{i}^{(j)}$, and $G_{j}=\frac{1}{N}X^{(j)\prime
}P_{Z^{(j)}}X^{(j)}$.
Denote
$
R=(1,1,1,1)^{\prime }\otimes \left(
\begin{array}{c}
\frac{1}{4}I_{k_{F}} \\
0_{k_{v},k_{F}}%
\end{array}%
\right) ,
$
a $4k\times k_{F}$ matrix. Then,
\begin{equation*}
\widehat{\Sigma }_{4S}=\frac{1}{N}R^{\prime }G^{-1}\widehat{\Sigma }_{0}G^{-1}R+\frac{1}{T}\widehat{\Omega }_{F},
\end{equation*}
where $\widehat{\Omega }_{F}$ is a consistent estimator of the long-run
variance of $F_t$.
\end{itemize}

\subsection{Consistency of the four-split estimator}
\label{subsection- asymptotics of our est}
\begin{theorem}
\label{theorem: 4split works} Assume that the samples $\{r_{it},$ $i=1,...,N,$
$t=1,...T\}$ and $\{F_{t},$ $t=1,...,T\}$ come from a data-generating
process that satisfies  factor pricing model (\ref{eq: main formulation of
dgp}) and assumptions FACTORS, LOADINGS and ERRORS. Let both $N$ and $T$ increase to infinity, then
\begin{shrinkeq}{-2ex}
$$\sqrt{T}(\widehat{\lambda}_{4S,1}-\lambda_1)=\sqrt{T}(\widetilde{\lambda}_{1}-\lambda_1)+O_p(1/\sqrt{N})\Rightarrow N(0,\Omega_F),$$\end{shrinkeq}
$$
\sqrt{\min\{N,T\}}(\widehat{\lambda}_{4S,2}-\lambda_2)=O_p(1).
$$
\end{theorem}

Theorem \ref{theorem: 4split works} establishes the consistency rate for the four-split estimator $\widehat{\lambda}_{4S}$ under exactly the same assumptions we showed the failure of the two-pass procedure. The four-split estimator for the risk premia on the strong observed factor is $\sqrt{T}$-consistent, asymptotically equivalent to $\widetilde\lambda_1$ and asymptotically gaussian, while the four-split estimate of the risk premia on the weak observed factor is consistent, and the rate of convergence depends on the relative size of $N$ and $T$. Theorem \ref{theorem: 4split works} shows that the four-split estimator has superior asymptotic properties in comparison to the classical two-pass procedure.

\textbf{Remark.} If we consider an asymptotic setting where $N\to\infty$, while $T$ is fixed, we can prove that under  slightly stronger assumptions on the error terms, the four-split estimate is consistent for ex-post risk premia $\widetilde\lambda$, and the $t$-statistics are asymptotically gaussian. The distinction between ex-post and ex-ante risk premia is discussed in Shanken (1992) as well as in Raponi, Robotti and Zaffaroni (2017) and in Kim and Skoulakis (2018). We need to strengthen the cross-sectional dependence assumptions on the error terms to the extent that the Law of Large Numbers and Cental Limit Theorems hold when summation is done over the cross-sectional index only. Assumptions in Raponi, Robotti and Zaffaroni (2017) and Kim and Skoulakis (2018) are of this type. The assumptions in Theorem 2 are somewhat weaker than needed for fixed $N$ setting as getting limit theorems when both summation indexes increase to infinity is easier, especially given the over-time independence of error-terms.

\textbf{Remark.} This papers considers unconditional pricing models only, but the estimation procedure can be adapted to work in some conditional pricing applications as well. In conditional pricing models we have either exposure coefficients $\beta_i$'s or risk premia $\lambda$ (or both) that are not constant but change slowly  over time. By using as a dependent variable  the average of the excess returns over a narrow window of the most recent observations, by using only $\beta_i^{(4)}$ as the regressor on the IV stage, and by adjusting the size of the most recent subsample $T_4$ to guarantee that $\beta_i$ is nearly stable, we can end up with a valid estimate of the current risk premia. Developing the details of such estimation is left for  future research.

\textbf{Remark.} One important assumption for the validity of our procedure is that we know the number of missing factors $k_v$. One may compose our estimator with a consistent selector of the number of factors  as Onatski (2009), Bai and Ng (2002) or Gagliardini, Ossola and Scaillet (2016b) do.

\section{Inference procedures using four-split estimator} \label{section- inference}

Theorem \ref{theorem: 4split works}  shows that the new four-split estimator is consistent but does not provide a basis for confidence set construction or testing. In order to make use of Theorem \ref{theorem: 4split works}    the researcher must know which observed factors are strong, and with that knowledge s/he can construct a confidence set for the risk premia only on the strong observed factor.  Apparently, the stated assumptions are not strong enough to obtain the asymptotic distribution of the full four-split estimator.  Below we formulate the needed additional high-level assumptions and establish a result about statistical inferences using the four-split estimator. We also provide primitive assumptions that will guarantee  the validity of the additional assumptions  in  examples.

For a set of vectors $a_j$, we denote by $(a_j)_{j=1}^4=(a_1^\prime,...,a_4^\prime)^\prime$ a long vector consisting of the four vectors stacked upon each other; we denote by $(a_{jj^*})_{j<j^*}$ the vectors $a_{jj^*}$ stacked together.

\noindent\textbf{Assumption GAUSSIANITY}
Assume that the following convergence holds:
\begin{align*}
 \frac{1}{\sqrt{N}}\sum_{i=1}^N
\left(
  \begin{array}{c}
    \sqrt{T}\gamma_i\overline{e}_i\\
    (\sqrt{T}\gamma_iu_i^{(j)})_{j=1}^4 \\
    (T\overline{e}_iu_i^{(j)})_{j=1}^4\\
    (Tu_i^{(j)}u_i^{(j^*)})_{j<j^*}
  \end{array}
\right)=\frac{1}{\sqrt{N}}\sum_{i=1}^N\xi_i&\Rightarrow \xi=
\left(
  \begin{array}{c}
    \xi_{\gamma e} \\
    (\xi_{\gamma j})_{j=1}^4 \\
    (\xi_{ej})_{j=1}^4 \\
    (\xi_{j,j^*})_{j<j^*} \\
  \end{array}
\right),
\end{align*}
where $\xi$ is a gaussian  vector with mean zero and covariance  $\Sigma_\xi$.

\medskip

\noindent\textbf{Assumption COVARIANCE} Assume that
$
\frac{1}{N}\sum_{i=1}^N\xi_i\xi_i^\prime\to^p \Sigma_\xi,
$
where $\xi_i$ and $\Sigma_\xi$ are defined in Assumption GAUSSIANITY.

\medskip

The assumptions we maintained in the previous sections are enough to guarantee that $\frac{1}{\sqrt{N}}\sum_{i=1}^N\xi_i$ is $O_p(1)$. Assumption GAUSSIANITY establishes the asymptotic distribution of that quantity, while Assumption COVARIANCE allows one to construct valid standard errors. Below we provide sufficient conditions for the two new assumptions in the two leading examples discussed before: one where the observed factors are independent from the errors and the example of factor-driven conditional heteroskedasticity.
\begin{lemma}
\label{lem: CLT for ERRORS^*} Assume that Assumption ERROR$^*$ holds and additionally,
\begin{itemize}
\item[(i)] $E\|F_t\|^8<\infty$; $E\|\frac{1}{|T_j|}\sum_{t\in T_j}F_tF_t^\prime-\Sigma_F\|\to 0$;
\item[(ii)] $\max_i\|\gamma_i\|<C;$
\item[(iii)] $\frac{1}{N}\mathrm{tr}(\mathcal{E}_{N,T}^2)\to a_2$ and $\frac{1}{N}\gamma^\prime\mathcal{E}_{N,T}\gamma\rightarrow \Gamma_{\sigma
}$, where $\Gamma_{\sigma
}$ is a full rank matrix;
\item[(iv)] $
\frac{1}{N^2}\sum_{i_1=1}^N\sum_{i_2=1}^N\sum_{i_3=1}^N \sum_{i_4=1}^N\left|Ee_{i_1t}e_{i_2t}e_{i_3t}e_{i_4t}\right|<C;
$
\end{itemize}
then  Assumption GAUSSIANITY holds. If in addition
\begin{shrinkeq}{-2ex}
$$
\|\mathcal{E}_{N,T}-\mathrm{dg}(\mathcal{E}_{N,T})\|\to 0 ~\mbox{ as } N,T\to\infty,
$$
\end{shrinkeq}%
then Assumption COVARIANCE holds as well.
\end{lemma}

\begin{lemma}
\label{lem: CLT for example 3} Assume we have a setting as in Example
3. Assume additionally that conditions (i) and (ii) of Lemma \ref{lem: CLT for ERRORS^*} hold and the following is true:
\begin{itemize}
\item[(i)] $E\left[(\|F_t\|^8+1)\|w_t\|^8\right]<\infty$;
\item[(ii)] $\frac{1}{N}\sum_{i=1}^N\sigma_i^4\to\mu_4$ and  $\frac{1}{N}\sum_{i=1}^N\sigma_i^2\gamma_i\gamma_i^\prime\to\Gamma_\sigma$, where $\Gamma_\sigma$ is a full rank matrix.
\end{itemize}
Then  Assumption GAUSSIANITY holds. If in addition  $\Gamma_{\pi\gamma}=0$, then Assumption COVARIANCE holds as well.
\end{lemma}

Assumption GAUSSIANITY is a result of strengthening moment restrictions (condition (i) in both Lemmas), guaranteeing that the asymptotic covariance matrix is well defined and full rank (condition (iii) in Lemma \ref{lem: CLT for ERRORS^*} and condition (ii) in Lemma \ref{lem: CLT for example 3}) and further restricting cross-sectional dependence (condition (iv) in Lemma \ref{lem: CLT for ERRORS^*}).

From a theoretical perspective, the derivation of a proper Central Limit Theorem in a factor model setting with a relatively free cross-sectional dependence structure is a major endeavor for two reasons. The first difficulty here is that the quite unrestrictive structure of the cross-sectional dependence of idiosyncratic error terms $e_{it}$ makes $\xi_i$ cross-sectionally dependent, though the correlation between $\xi_i$ and $\xi_{i^*}$ for $i\neq i^*$ converges to zero for large sample sizes. Without  imposing further discipline on the structure of dependence, it is hard to obtain a Central Limit Theorem.  Secondly, the components $\xi_{ej}$ and $\xi_{j,j^*}$ are quadratic forms in the initial errors. Here we use the asymptotic results established for exactly this setting in a separate paper by  Anatolyev and Mikusheva (2018). There, we exploit time-series conditional independence of errors to obtain a Central Limit Theorem for cross-sectional sums.

\begin{theorem}
\label{theorem: asymptotics of 4 split} Assume that the samples $\{r_{it},$ $i=1,...,N,$
$t=1,...T\}$ and $\{F_{t},$ $t=1,...,T\}$ come from a data-generating
process that satisfies  factor pricing model (\ref{eq: main formulation of
dgp}) and Assumptions FACTORS, LOADINGS, ERRORS and GAUSSIANITY as both $N$ and $T$ increase to infinity. Then $\sqrt{\min\{N,T\}}(\widehat{\lambda}_{4S,2}-\lambda_2)$ weakly converges to a mixed gaussian distribution. If in addition Assumption COVARIANCE holds, then
\begin{shrinkeq}{-2ex}
$$
\widehat\Sigma_{4S}^{-1/2}(\widehat{\lambda}_{4S}-\lambda)\Rightarrow N(0,I_k).
$$
\end{shrinkeq}
\end{theorem}

Theorem \ref{theorem: asymptotics of 4 split} suggests the use of $t$- and Wald statistics for the construction of confidence sets and  testing hypotheses about values of the risk premia. These inference procedures are standard and  can be implemented using standard econometrics software.

From a theoretical perspective, however, the asymptotics of the four-split estimator are not fully standard. The asymptotic distribution of the four-split estimator is not gaussian but rather  mixed gaussian, that is, the limit distribution of the four-split estimator can be written as a gaussian random vector with random variance.
To understand the intuition,  look at equation (\ref{eq:second step initial}) and notice that the
coefficient $\eta _{v,T}$ on the omitted variable $\mu _{i}$ is random, even
asymptotically.  This implies that the amount of information contained in the sample, which
is used to correct for the omitted-variable problem, is random as well, and
thus results in an asymptotically random covariance matrix.
Theorem \ref{theorem: asymptotics of 4 split} shows that  a properly constructed proxy for the asymptotic variance restores the asymptotic gaussianity of a multidimensional $t$-statistic  even when the estimator itself is not asymptotically gaussian. A similar phenomenon has appeared before in the literature on estimation of co-integrating relations.

Another important aspect of Theorem \ref{theorem: asymptotics of 4 split} is that inferences or construction of a proxy for the variance do not assume  knowledge of the number or identity of strong/weak factors. This is a desirable feature, as we do not have a procedure that can credibly differentiate between weak and strong factors.

As previously discussed, even though the main data set contains $NT$ observations, the risk premia cannot be estimated at a rate better than $\sqrt{T}.$ This can be seen from equation (\ref{eq: ideal cross sectional}), as even if we know the true values of $\beta_i$ the regression of $\overline{r}_i$ on $\beta_i$ has a true coefficient equal to $\widetilde{\lambda}=\lambda+\overline{F}-EF_t$. This means that the uncertainty associated with the deviations of $\frac{1}{T}\sum_{t=1}^TF_t$ from $EF_t$ is unavoidable. This also justifies the presence of the long-run variance of factors, $\Omega_F$, in the variance estimate $\widehat\Sigma_{4S}$. Theorem \ref{theorem: 4split works} also states that the difference between $\widehat\lambda_{4S}$ and $\widetilde\lambda$ is of order $\frac{1}{\sqrt{NT}}$. From the proof of Theorem \ref{theorem: asymptotics of 4 split} we  see that this difference is mixed gaussian, and the variance can be deduced from $\widehat\Sigma_{IV}$.
Typically, $\widetilde\lambda$ is infeasible. However, if all observed factors are portfolios themselves and are priced by the same model, then we have $\lambda=EF_t$. In such a case the literature suggests the use of an alternative feasible estimator $\widehat\lambda=\frac{1}{T}\sum_{t=1}^TF_t$, which in this case is equal to $\widetilde\lambda$. Thus, in this special case we have two competing estimators for $\lambda$ and can create a test for model specification. In particular, the statistic compares the difference between $\widehat\lambda_{4S}$ and $\widetilde\lambda$ to zero. The proof of Theorem \ref{theorem: asymptotics of 4 split} shows that $\widehat\lambda_{4S}-\widetilde\lambda$ converges to zero at the rate $\sqrt{NT}$, is asymptotically mixed gaussian, and $\widehat\Sigma_{IV}$ is a proper proxy for the variance that delivers a $\chi^2$ asymptotic distribution to the corresponding Wald statistic.

\section{Simulation evidence}

\label{section- simulations}

The goal of this section is to explore the size of potential deficiency of the two-pass and superior performance of the four-split procedure in a setting close to a real-life application. We expect  the largest effect to come from the omitted-variable bias due to missing factor structure. The size of the bias and the distortion of the t-test coverage  depend on the size of the missing factor, the strength of the  observed factors, and the correlation between observed and missing factors. We explore these relations below.

\subsection{Simulations using artificial data}

\textbf{Empirical setting.}
We calibrate the data-generating process in our simulations to match  the data set of the monthly returns on 100 Fama-French portfolios sorted by size and book-to-market and asset returns' relation with the 3 Fama-French factors (market, SmB, HmL). The data is taken from
Kenneth French's web-site. We  substitute missing values with zeros. The returns are value-weighted, the excess returns are calculated using one-month Treasury bills. The sample sizes are $%
N=100$ and $T=504$.

First, we run principle components (PC) on the panel of excess returns, call the first three PCs $G_{t}$
with their loadings $\gamma _{i},$ and the fourth main principle component $%
g_{t}$ with  loadings $\phi _{i}.$ We use the normalization $\sum_{t=1}^T(G_t',g_t)'(G_t',g_t)=I_4$,
making the variance of each factor $1/T$ (all results are invariant to the normalization).
We compute sample means of the loadings, $\mu _{\gamma }$ and $\mu _{\phi },$ and their
sample variances, $V_{\gamma }$ and $v_{\phi }$. We compute the residuals $%
\varepsilon _{it}$ left after four main PCs,  and compute their sample variance $\sigma
_{\varepsilon }^{2}.$  In order to preserve the relation between PCs and Fama-French factors,  we run a regression of $F_{t}$ on a constant and $G_{t},$ obtain intercepts $\eta _{0,F}$, slopes $\eta_F $ and residual variance matrix $\Sigma _{res}.$

The fraction of the total variation explained by
the four main PCs is approximately 73\%, 6\%, 3\% and 1\%. In order to have a proper comparison consistent with our theoretical results, we  measure the strength of a given factor as the total variation in the data it produces, that is, its strength is measured as the sum of squared loading on that factor times the variance of the factor. For example, for the fourth principle component $g_t$, its strength  is measured as $\frac{1}{T}\sum_{i=1}^N\phi_i^2$. The strengths of the four main PCs and Fama-French factors  are reported in Table \ref{table: observed strength}. These numbers can serve as a reference for the simulation results below.

\begin{table}[!htb]\centering
 \caption{The strength of principle components and Fama-French factors in the Fama-French data set, as measured by the total variation produced in the excess returns.}
 \vspace{3mm}
\begin{tabular}{l|cccc}
\hline\hline
        & 1st PC&  2nd PC  &  3rd PC   & 4th PC \\
\hline
Strength of principle components & 2816  & 239 & 113& 50 \\
\hline\hline
        & market&  SmB  &  HmL   & \\
\hline
Strength of Fama-French factors &2263  & 482 & 230 &\\
\hline\hline
\end{tabular}%
\label{table: observed strength}
\end{table}
\vspace{-5mm}

\paragraph{Simulation design.}
We simulate the data by following three steps. In the first step  we are trying to match the relation of the principle components and Fama-French factors. We simulate $G_{t}\sim i.i.d.N(0,I_{3}/T)$
and $\gamma _{i}\sim i.i.d.N(\mu _{\gamma },V_{\gamma })$, and then construct the simulated
`observed factors' by $F_{t}=\eta _{0,F}+\eta _{F}G_{t}+w_{t}$, where $%
w_{t}\sim i.i.d.N(0,\Sigma _{res})$.

As a second step we introduce one more factor to the returns, part of which will represent a missing factor structure in the errors, by simulating $%
g_{t}\sim i.i.d.N(0,1/T)$ and $\phi _{i}\sim i.i.d.$ $\vartheta _{\phi
}\cdot N(\mu _{\phi },v_{\phi })$. The parameter $\vartheta _{\phi }$
indexes the strength of the missing factor. Finally, we introduce one more observed factor, which
we label as $mom$ because we want to imitate the relation of the momentum factor to the PCs.
The analogy stops here: we do not claim to mimic the true momentum factor. We simulate $%
mom_{t}=\eta _{0,mom}+\eta _{mom}G_{t}+u_{t}+v_{t},$ where $\eta _{0,mom}$ and $\eta _{mom}$ are coefficients from a regression of the momentum factor on the PCs in the data.   The simulated error consists of two parts: $v_{t}\sim i.i.d.N(0,\varphi \sigma _{mom}^{2})$ is uncorrelated with returns or other factors, while $u_{t}\sim
i.i.d.N(0,(1-\varphi )\sigma _{mom}^{2})$  will appear as a part of excess returns and is the reason for using $mom$ in the pricing of assets. Here $\sigma _{mom}^{2}$ is the estimated variance of the residuals from a regression of $mom$ on PCs. In all simulations, we set $\varphi =0.001.$

We generate loadings on $u_{t}$ according to
$
\delta _{i}=\frac{\alpha \phi _{i}/\sqrt{T}+\xi _{i}}{\sqrt{(1-\varphi
)\sigma _{mom}^{2}}},
$
where $\xi _{i}\sim N(0,\sigma _{\xi }^{2}),$ so that they are (imperfectly)
correlated with the loadings on $g_t$. Here by increasing $\alpha$ we can increase correlation, and by increasing $\sigma _{\xi }^{2}$ we can increase the strength of $mom$.

For the third step we generate a cross-section of returns and impose a correct pricing model.
We generate the de-meaned part of excess returns according to
$
r_{it}^*=G_{t}\gamma _{i}+g_{t}\phi _{i}+u_{t}\delta
_{i}+\epsilon _{it},
$
where $\epsilon _{it}\sim i.i.d.N(0,\sigma _{\varepsilon }^{2})$ is an idiosyncratic error. The
implied true betas  are
\begin{align*}
\beta _{i} =\mathrm{var}\binom{F_{t}}{mom_{t}}^{-1}\mathrm{cov}\left( \binom{F_{t}}{mom_{t}}%
,r_{it}^*\right) =\mathrm{var}\binom{F_{t}}{mom_{t}}^{-1}\binom{\eta _{F}\gamma _{i}/T}{\eta
_{mom}\gamma _{i}/T+(1-\varphi )\sigma _{mom}^{2}\delta _{i}}.
\end{align*}
Correctly priced excess returns are obtained by adding risk premia:
$r_{it}=r^*_{it}+\lambda \beta _{i}$, where we set $\lambda $
to be sample means of the Fama-French and momentum factors from the data.

\paragraph{Simulation results.}
In all simulation experiments, we read off the biases, `absolute' biases,
standard deviations of the estimates for  momentum risk premium, as well as the actual
rejection rates of the null that the risk premium on the momentum factor
equals its true value using the 5\% nominal critical values. As we saw in equation (\ref{eq:second step initial}) the omitted bias depends on the random variable $\eta_v$, which in different realizations of factors may be positive or negative with equal probability and washes out in repeated draws
of time-series processes. The `absolute' bias is actually a better characterization of the centrality of a distribution.  It is an absolute value of the bias averaged across $R_{(i)}=100$ draws of simulated cross sections (such as $\gamma _{i},$ $\phi _{i},$ etc.) averaged
across each of $R_{(t)}=100$ draws of time-series processes (such as $G_{t},$
$F_{t},$ $mom_{t}$, etc.). Thus the total number of simulations is $%
R_{(i,t)}=R_{(i)}R_{(t)}=10,000$.

\begin{figure}[!htb]
    \includegraphics[scale=0.7]{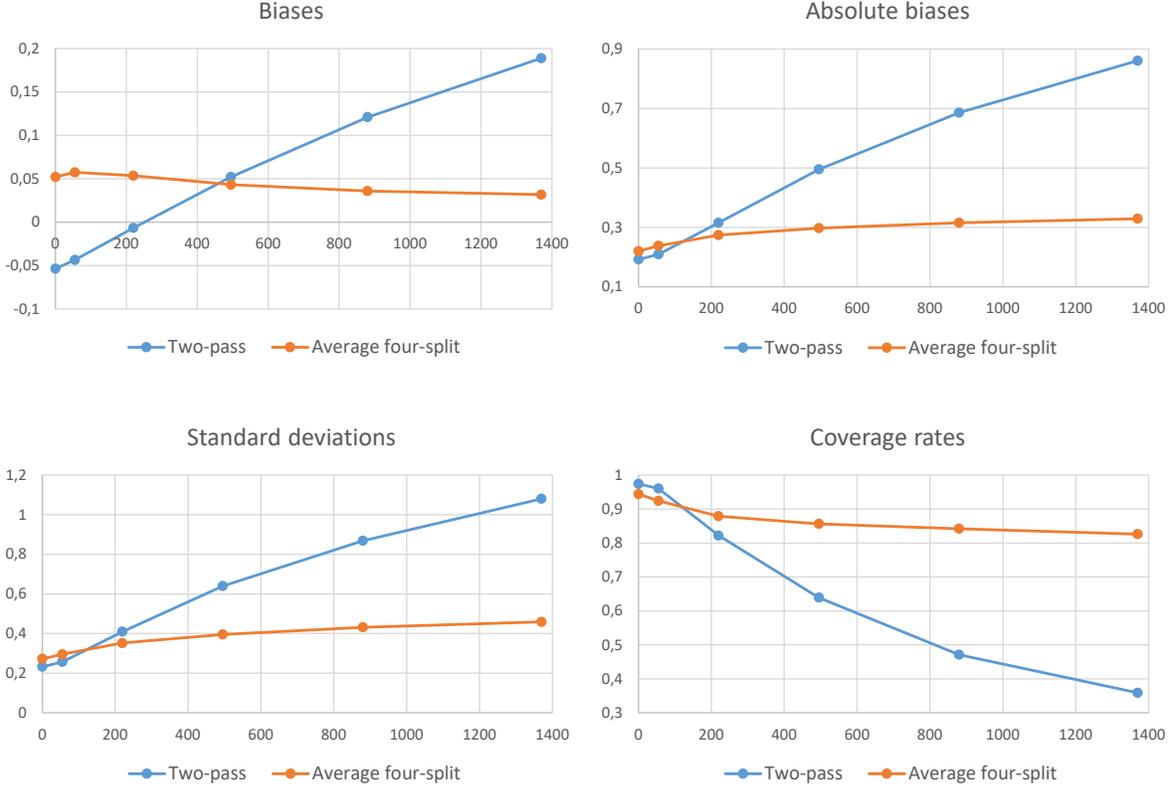}
    \caption{Simulated bias, absolute bias and standard deviation of the estimate  and coverage of the $t$-statistics confidence for the risk premia on $mom$ for two-pass and four-split methods. The strength of factor $g_t$ measured as total variation explained is on the horizontal axis. Number of simulations of the time series process $R_{(t)}=100$. For each time series draw, we simulate $R_{(i)}=100$ cross-section draws. Total number of simulations is $10,000.$ }%
\label{figure- experiment 1}
\end{figure}

In the first set of experiments, we explore the effect of a change in the strength of the missing factor. We set $\alpha =0.1$ and $\sigma _{\xi
}^{2}=0.3,$ and then vary $\vartheta _{\phi }$ from 0 to 5, the value 0
meaning no missing factors, the value 1 corresponding to the strength of the fourth PC. Figure \ref{figure- experiment 1} shows the performance measures for the
conventional two-pass and proposed average four-split estimators, with the
strength of the missing factor (as measured by the total variation explained by the factor, $\frac{1}{T}\sum_{i=1}^{N}\phi
_{i}^{2}$) on the horizontal axis. From the bias panel, one can see
that the two-pass estimator is unpredictable in `biasedness' properties, and
its bias can grow indefinitely, while the four-split estimator is much more
stable, with the bias staying bounded while the missing factor grows in
strength. More importantly, though, is that the absolute bias of the two-pass
estimator grows fewfold larger than that of the four-split estimator. About
the same happens to these estimators' standard deviations: the gap in
variability grows fewfold as the missing factor grows in strength. Finally,
the test size distortions may be smallest for the two-pass estimator in the
absence of missing factors; they quickly become huge when the missing factor
increases in importance. One can see that if the missing factor is of the size of SmB (see Table \ref{table: observed strength})  one can easily get coverage around 60\% instead of the declared 95\%.

\begin{figure}[!htb]
    \includegraphics[scale=0.7]{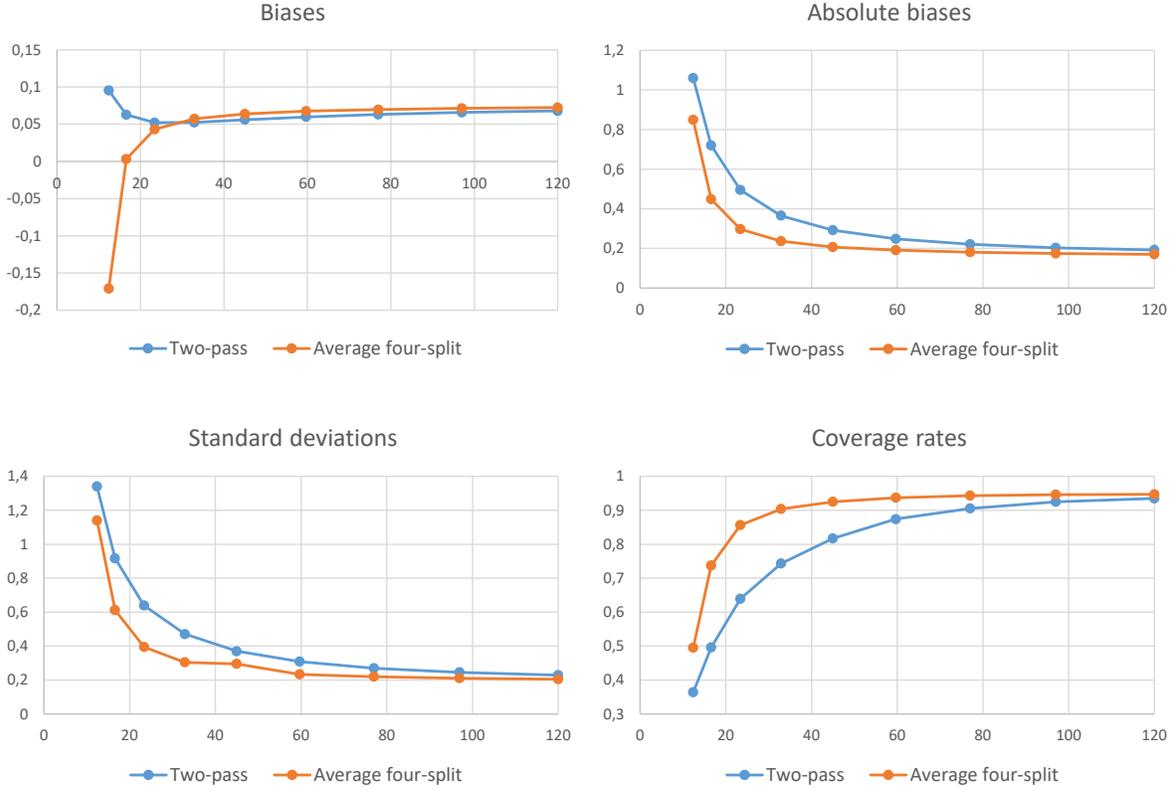}
    \caption{Simulated bias, absolute bias and standard deviation of the estimate  and coverage of the $t$-statistics confidence for the risk premia on $mom$ for two-pass and four-split methods. The strength of factor $mom$  is on the horizontal axis. Number of simulations of the time series process $R_{(t)}=100$. For each time-series draw, we simulate $R_{(i)}=100$ cross-section draws. Total number of simulations is $10,000.$ }%
\label{figure- experiment2}
\end{figure}

In the second set of experiments, we set $\vartheta _{\phi }=3$ and $\alpha
=0.1,$ and then vary $\sigma _{\xi }^{2}$ from 0.1 to 0.9 with a step of
0.1, which makes the strength of the missing factor (as measured by $\frac{1}{T}
\sum_{i=1}^{N}\phi _{i}^{2}$) fixed at 495, and the strength of
the momentum factor (as measured by $\sum_{i=1}^{N}\beta _{i,4}^{2}\frac{1}{T}\sum_{t=1}^{T}\left( mom_{t}-\overline{mom}\right) ^{2}$)  increase
from 12.4 to 120.0. At the same time, the correlation between the missing
and momentum factors (as measured by the sample correlation between $\phi
_{i}$ and $\beta _{i,4}$)  decreases from 0.917 to 0.336. Figure \ref{figure- experiment2} shows
the performance measures for the conventional two-pass and proposed average
four-split estimators, with the strength of the momentum factor on the
horizontal axis. While the ranking by the signed bias may not be
predictable, there is clear dominance of the four-split estimator in terms
of  absolute bias, standard deviation, and actual rejection rates. The
latter may be quite far from the nominal size when the momentum factor exhibits
too much weakness, which in this case would be violation of Assumption LOADINGS.

\subsection{Size of the effect in an empirical application}

\label{section- application}

In this subsection we run the two-pass and four-split procedures on the data set of the monthly returns on 100 Fama-French portfolios sorted by size and book-to-market using  3 Fama-French factors (market, SmB, HmL) and the momentum factor as observed factors. The data are described previously. We estimate the model with the Fama-French factors only and then  with all  four factors. We report the results  in Table \ref{table: FFapplication}, where for estimation of the long-run variance of observed factors we use the Newey-West estimator with 4 lags (though the results are not sensitive to a lag choice).

 \vspace{-3mm}

\begin{table}[!htb]\centering
 \caption{Estimates of the risk premia on the Fama-French factors and the momentum factor and tests of specification, using monthly returns on 100 portfolios sorted by size and book-to-market. The covariance matrix is computed using the Newey-West estimator with 4 lags. The sample size is $T=504$.  }
 \vspace{3mm}
\begin{tabular}{l|cccc|cc}
\hline\hline
 & Market & SMB & HML & MOM & Wald & p-val\\
 \hline
average ex. return & $\underset{0.216}{0.527}$ & $\underset{0.138}{0.187}$ & $\underset{0.152}{0.401}$ & $\underset{0.205}{0.708}$ &&\\
\hline\hline
\multicolumn{7}{c}{Model with 3 Fama-French factors}\\
\hline\hline
Two-pass & $\underset{0.218}{0.489}$ & $\underset{0.146}{0.185}$ & $\underset%
{0.165}{0.440}$&&2.25&0.522 \\
Four-split & $\underset{0.219}{0.510}$ & $\underset{0.152}{0.167}$ & $%
\underset{0.165}{0.439}$&&1.08&0.782\\
\hline\hline
\multicolumn{7}{c}{Model with 3 Fama-French factors and momentum}\\
\hline\hline
Two-pass & $\underset{0.218}{0.584}$ & $\underset{0.145}{0.162}$ & $\underset%
{0.161}{0.461}$ & $\underset{0.385}{1.860}$ &21.99& 0.000\\
Four-split & $\underset{0.223}{0.534}$ & $\underset{0.159}{0.210}$ & $%
\underset{0.164}{0.436}$ & $\underset{0.637}{0.542}$&3.96&0.411\\
\hline\hline
\end{tabular}%
\label{table: FFapplication}
\end{table}
 \vspace{-3mm}

Since all the observed factors are tradable, we have an alternative (and the most efficient) estimator of the risk premia, $\widehat\lambda=\frac{1}{T}\sum_{t=1}^TF_t$, the average excess return, which is not available for non-tradable factors. This allows us to discuss the quality of different estimates relative to this benchmark. Also, having two valid estimates with different efficiency allows us to test for a correct specification of the linear pricing model (the J-test). The specification test based on the Wald statistic is equal to the squared difference between the (two-pass or four-split) estimate and the average factor, weighted by an inverse of the difference in covariance matrices. The validity of such a test comes from the proof of Theorem \ref{theorem: asymptotics of 4 split}. The results of the specification tests and the corresponding p-values appear in Table \ref{table: FFapplication}.

Notice that when we estimate the model with the three Fama-French factors only, both two-pass and four-split estimates are statistically indistinguishable from the average returns, so the Wald tests fail to reject the correct specification of the pricing model. Also, there is almost no cost in terms of efficiency of either estimate. When, however, we estimate the model with all four factors, the two-pass procedure produces a very high value for the risk premia  on the momentum, and also falsely overstates the accuracy of that estimate to such an extent that the two-pass procedure strongly rejects the linear pricing model with four factors. This is an unfortunate outcome since the correct specification of the three-factor model implies a correct specification of the pricing model whenever one more factor is added. We attribute this strange behavior of the two-pass procedure to the momentum factor having only weak correlation with the returns. The four-split procedure, however, produces an estimate of the momentum risk premium close to the average excess return and accepts the correctness of the specification. The four-split estimate has much larger standard errors in comparison to the average excess returns, which is an implicit confirmation of the weakness of the momentum factor.

\section*{References}

\textsc{Anatolyev, S.} (2012): \textquotedblleft Inference in Regression Models with
Many Regressors,\textquotedblright\ \textit{Journal of Econometrics},
170(2), 368-382.

\noindent \textsc{Anatolyev, S. and Mikusheva, A.} (2018): \textquotedblleft Limit Theorems for Factor Models,\textquotedblright\ unpublished manuscript available at \texttt{http://economics.mit.edu/files/15221}

\noindent \textsc{Andrews, D.W.K.} (2005): \textquotedblleft Cross-Section Regression with
Common Shocks,\textquotedblright\ \textit{Econometrica}, 73(5), 1551-1585.

\noindent \textsc{Angrist, J.D., G.W. Imbens and A. Krueger} (1999): \textquotedblleft
Jackknife Instrumental Variables Estimation,\textquotedblright\ \textit{%
Journal of Applied Econometrics}, 14, 57-67.

\noindent \textsc{Bai, J. and Ng, S.} (2002): \textquotedblleft Determining the Number of
Factors in Approximate Factor Models,\textquotedblright\ \textit{Econometrica%
}, 70, 191-221.

\noindent \textsc{Bai, J. and S. Ng} (2006): \textquotedblleft Confidence Intervals for
Diffusion Index Forecast and Inference with Factor-Augmented
Regressions\textquotedblright , \textit{Econometrica}, 74, 1133-1155.

\noindent \textsc{Bekker, P.A.} (1994): \textquotedblleft Alternative Approximations to the
Distributions of Instrumental Variable Estimators,\textquotedblright\
\textit{Econometrica}, 62, 657-681.

\noindent \textsc{Bryzgalova S.} (2016): \textquotedblleft Spurious Factors in Linear Asset
Pricing Models\textquotedblright, manuscript, Stanford Graduate School of Business.

\noindent \textsc{Burnside, C.} (2016): \textquotedblleft Identification and Inference in
Linear Stochastic Discount Factor Models with Excess
Returns,\textquotedblright\ \textit{Journal of Financial Econometrics},
14(2), 295-330.

\noindent \textsc{Cochrane, J.} (2001): \textit{Asset Pricing}, Princeton University Press.

\noindent \textsc{Dufour, J.M. and  Jasiak, J.} (2001): \textquotedblleft Finite Sample Limited Information Inference Methods for Structural Equations and Models With Generated Regressors,\textquotedblright\ \textit{International Economic Review}, 42, 815-844.

\noindent \textsc{Fama, E.F. and French, K.R.} (1993): \textquotedblleft Common Risk Factors in
the Returns on Stocks and Bonds,\textquotedblright\ \textit{Journal of
Financial Economics}, 33(1), 3-56.

\noindent \textsc{Fama, E.F. and MacBeth, J.} (1973): \textquotedblleft Risk, Return and
Equilibrium: Empirical Tests,\textquotedblright\ \textit{Journal of
Political Economy}, 81, 607-636.

\noindent \textsc{Feng, G., Giglio, S. and Xiu, D.} (2019): \textquotedblleft Taming the Factor Zoo: A Test of New Factors,\textquotedblright\ NBER Working Paper No. 25481.

\noindent \textsc{Gagliardini, P., Ossola, E. and Scaillet, O.} (2016a) \textquotedblleft
Time-Varying Risk Premium in Large Cross-Sectional Equity Data
Sets,\textquotedblright\ \textit{Econometrica}, 86, 985-1046.

\noindent \textsc{Gagliardini, P., Ossola, E. and Scaillet, O.} (2016b) \textquotedblleft
A Diagnostic Criterion for Approximate Factor Structure,\textquotedblright\ \textit{Swiss Finance Institute Research Paper}, 16-51.

\noindent \textsc{Giglio, S. and Xiu, D.} (2017): \textquotedblleft Inference on Risk Premia in the Presence of Omitted Factors,\textquotedblright\ NBER Working Paper No. 23527.

\noindent \textsc{Gospodinov, N., Kan, R. and Robotti, C.} (2017): \textquotedblleft Spurious
Inference in Reduced-Rank Asset-Pricing Models,\textquotedblright\
\textit{Econometrica}, 85(5), 1613-1628.

\noindent \textsc{Hansen, C., Hausman, J. and Newey, W.K.} (2008): \textquotedblleft Estimation
with Many Instrumental Variables,\textquotedblright\ \textit{Journal of
Business \& Economics Statistics}, 26, 398-422.

\noindent \textsc{Harvey, C.R., Liu, Y. and Zhu, H.} (2016): \textquotedblleft ... and the
Cross-Section of Expected Returns,\textquotedblright\ \textit{Review of
Financial Studies}, 29(1), 5-68.

\noindent \textsc{Heyde, C. and Brown, B.} (1970): \textquotedblleft On the Departure from
Normality of a Certain Class of Martingales,\textquotedblright\ \textit{%
Annals of Mathematical Statistics}, 41(6), 2161-2165.

\noindent \textsc{Jagannathan, R. and Wang, Z.} (1996) The Conditional CAPM and the
Cross-Section of Expected Returns. \textit{Journal of Finance}, 51(1), 3-53.

\noindent \textsc{Jegadeesh, N., Noh, J., Pukthuanthong, K., Roll, R., and Wang, J.} (2019): \textquotedblleft Empirical Tests of Asset Pricing Models with Individual Assets: Resolving the Error-in-Variables Bias in Risk Premium Estimation,\textquotedblright\ \textit{Journal of Financial Economics}, forthcoming.

\noindent \textsc{Jegadeesh, N. and Titman, S.} (1993): \textquotedblleft Returns to Buying
Winners and Selling Losers: Implications for Stock Market
Efficiency,\textquotedblright\ \textit{Journal of Finance}, 48(1), 65-91.

\noindent \textsc{de Jong, P.} (1987): \textquotedblleft A Central Limit Theorem for
Generalized Quadratic Forms,\textquotedblright\ \textit{Probability Theory
and Related Fields}, 75, 261-277.

\noindent \textsc{Kan, R. and Zhang, C.} (1999): \textquotedblleft Two-Pass Tests of Asset
Pricing Models with Useless Factors,\textquotedblright\ \textit{Journal of
Finance}, 54, 203-235.

\noindent \textsc{Kim, S. and Skoulakis, G.} (2018): \textquotedblleft Ex-post risk premia estimation and asset pricing tests using large
cross sections: The regression-calibration approach,\textquotedblright\ \textit{Journal of
Econometrics}, 24, 159-188.

\noindent \textsc{Kleibergen, F.} (2009): \textquotedblleft Tests of Risk Premia in Linear
Factor Models,\textquotedblright\ \textit{Journal of Econometrics}, 149(2),
149-173.

\noindent \textsc{Kleibergen, F. and Zhan, Z.} (2015): \textquotedblleft Unexplained Factors and
Their Effects on Second Pass R-squared's,\textquotedblright\ \textit{Journal
of Econometrics}, 189, 101-116.

\noindent \textsc{Kozak, S., Nagel, S. and Santosh, S.} (2018): \textquotedblleft Shrinking the Cross-Section,\textquotedblright\ \textit{Journal of Financial Economics}, forthcoming.

\noindent \textsc{Lettau, M. and Ludvigson, S.} (2001): \textquotedblleft Resurrecting the
(C)CAPM: A Cross-Sectional Test When Risk Premia Are
Time-Varying,\textquotedblright\ \textit{Journal of Political Economy},
109(6), 1238-1287.

\noindent \textsc{Lettau, M. and Pelger, M.} (2018): \textquotedblleft  Estimating Latent Asset-Pricing Factors,\textquotedblright\ NBER Working Paper No. 24618.

\noindent \textsc{Lintner, J.} (1965): \textquotedblleft Security Prices, Risk, and Maximal
Gains from Diversification,\textquotedblright\ \textit{Journal of Finance},
20, 587-615.

\noindent \textsc{Lewellen, J., Nagel, S., and Shanken, J.} (2010): \textquotedblleft A Skeptical Appraisal of Asset Pricing Tests,\textquotedblright\ \textit{Journal of Financial Economics},
96, 175-194.

\noindent \textsc{Newey, W.K. and  Windmeijer, F.} (2009): \textquotedblleft Generalized Method of Moments With Many Weak Moment Conditions,\textquotedblright\ \textit{Econometrica}, 77, 687-719.

\noindent \textsc{Onatski, A.} (2012): \textquotedblleft Asymptotics of the Principal
Components Estimator of Large Factor Models with Weakly Influential
Factors,\textquotedblright\ \textit{Journal of Econometrics}, 168, 244-258.

\noindent \textsc{P\'{a}astor, L. and Stambaugh, R.F.} (2003): Liquidity risk and expected stock returns,\textquotedblright\ \textit{Journal of Political Economy}, 111(3), 642-685.

\noindent \textsc{Raponi, V., Robotti, C., and Zaffaroni, P.} (2017): \textquotedblleft Testing Beta-Pricing Models Using Large Cross-Sections,\textquotedblright\ Working paper, Yale University.

\noindent \textsc{Shanken, J.} (1992): \textquotedblleft On the Estimation of Beta-Pricing
Models,\textquotedblright\ \textit{Review of Financial Studies}, 5, 1-33.

\noindent \textsc{Sharpe, W.F.} (1964): \textquotedblleft Capital Asset Prices: A Theory of
Market Equilibrium under Conditions of Risk,\textquotedblright\ \textit{%
Journal of Finance}, 19, 425-442.

\noindent \textsc{Staiger, D. and Stock, J.H.} (1997): \textquotedblleft Instrumental Variables
Regression with Weak Instruments,\textquotedblright\ \textit{Econometrica},
65(3), 557-586.

 \section*{Appendix A: proofs}
\label{section- proofs}

Note that Lemma S1 often referred to is contained in the Supplementary Appendix.

\paragraph{Proof of Theorem \protect\ref{theorem: FMB fails}.}

Assumption FACTORS guarantees that  $\sqrt{T}(\widetilde\lambda-\lambda)%
\Rightarrow N(0,\Omega_F)$.
The first-pass (time series) regression yields equation (\ref{eq: first stage estimate}),
where we use Assumption FACTORS, and $o_{p}(1)$ appears from the
difference between $\Sigma _{F}$
and $T^{-1}\sum_{t=1}^{T}\widetilde{F}_{t}\widetilde{F}_{t}^{\prime }$.

Denote $Q_{T}=\left(
\begin{array}{cc}
I_{k_{1}} & 0_{k_{1},k_{2}} \\
0_{k_{2},k_{1}} & \sqrt{T}I_{k_{2}}%
\end{array}%
\right) .$ Notice that $Q_{T}/\sqrt{T}\rightarrow \mathcal{I}_{k_{2}}$. Below we prove the following statement: as $N,T\rightarrow \infty $,
\begin{shrinkeq}{-2ex}
\begin{equation}
\frac{1}{N}\sum_{i=1}^{N}Q_{T}\widehat{\beta }_{i}\widehat{\beta }_{i}^{\prime
}Q_{T}\Rightarrow (I_{k_{\beta }};\widetilde{\eta })\Gamma
(I_{k_{\beta }};\widetilde{\eta })^{\prime }+\mathcal{I}_{k_{2}}\Sigma _{u}%
\mathcal{I}_{k_{2}}.  \label{eq: asympt for denominator}
\end{equation}
\end{shrinkeq}%
Indeed,
\begin{shrinkeq}{-2ex}
\begin{align}
\frac{1}{N}\sum_{i=1}^{N}Q_{T}\widehat{\beta }_{i}\widehat{\beta }%
_{i}^{\prime }Q_{T}& =\frac{1}{N}\sum_{i=1}^{N}\left( Q_{T}\beta _{i}+Q_{T}%
\frac{\eta _{T}}{\sqrt{T}}\mu _{i}+Q_{T}u_{i}\right) \left( Q_{T}\beta
_{i}+Q_{T}\frac{\eta _{T}}{\sqrt{T}}\mu _{i}+Q_{T}u_{i}\right) ^{\prime }
\notag \\
& =\frac{1}{N}\sum_{i=1}^{N}\left( (I_{k_{\beta }};\widetilde{\eta }%
_{T})\gamma _{i}+Q_{T}u_{i}\right) \left( (I_{k_{\beta }};\widetilde{\eta }%
_{T})\gamma _{i}+Q_{T}u_{i}\right) ^{\prime },  \label{eq: in between denom}
\end{align}
\end{shrinkeq}%
where $\widetilde{\eta }_{T}=Q_{T}\eta _{T}/\sqrt{T}\Rightarrow \mathcal{I}%
_{k_{2}}\eta =\widetilde{\eta }$ is $k_{F}\times k_{v}$ gaussian random
matrix. Let us show that
\begin{shrinkeq}{-1ex}
\begin{equation}
\frac{T}{N}\sum_{i=1}^Nu_{i}u_{i}^{\prime }\rightarrow \Sigma _{u}.
\label{eq: variance of error}
\end{equation}
\end{shrinkeq}%
Indeed, due to statement (4) of Lemma S1 we have that
$
\frac{1}{TN}\sum_{i=1}^{N}\sum_{t=1}^{T}\sum_{s=1,s\neq t}^{T}\widetilde{F}%
_{t}\widetilde{F}_{s}^{\prime }e_{it}e_{is}=o_{p}(1).
$
Thus,
\begin{align*}
\frac{T}{N}\sum_{i=1}^Nu_{i}u_{i}^{\prime }& =\Sigma _{F}^{-1}\left( \frac{1}{TN}%
\sum_{i=1}^{N}\sum_{t=1}^{T}\sum_{s=1}^{T}\widetilde{F}_{t}\widetilde{F}%
_{s}^{\prime }e_{it}e_{is}\right) \Sigma _{F}^{-1} \\
& =\Sigma _{F}^{-1}\left( \frac{1}{TN}\sum_{i=1}^{N}\sum_{t=1}^{T}\widetilde{%
F}_{t}\widetilde{F}_{t}^{\prime }e_{it}^{2}\right) \Sigma
_{F}^{-1}+o_{p}(1)\rightarrow ^{p}\Sigma _{u},
\end{align*}%
where the last convergence comes from statement (3) of Lemma S1. Statement (5) of
Lemma S1 implies
\begin{equation}
\frac{\sqrt{T}}{N}\sum_{i=1}^{N}\gamma _{i}u_{i}^{\prime }\rightarrow
^{p}0_{k,k_{F}}.  \label{eq: something}
\end{equation}%
Combination of equations (\ref{eq: in between denom})--(\ref{eq: something}) and
Assumption LOADINGS implies (\ref{eq: asympt
for denominator}).

For the ``attenuation bias,''
\begin{shrinkeq}{-1ex}
\begin{equation*}
\left(
\begin{array}{c}
\sqrt{T}B^A_{1} \\
B^A_{2} \\
\end{array}%
\right) =Q_{T}^{-1}\sqrt{T}B^A=-\left( \frac{1}{N}Q_{T}\sum_{i=1}^N\widehat{\beta
}_{i}\widehat{\beta }_{i}^{\prime }Q_{T}\right) ^{-1}\frac{Q_{T}}{\sqrt{T}}%
\frac{T}{N}\sum_{i=1}^Nu_{i}u_{i}^{\prime }\widetilde{\lambda }.
\end{equation*}
\end{shrinkeq}
Combining equations (\ref{eq: asympt for denominator}), (\ref{eq: variance
of error}), $\widetilde{\lambda }\rightarrow ^{p}\lambda $ and $Q_{T}/\sqrt{T%
}\rightarrow \mathcal{I}_{k_{2}},$ we arrive at
\begin{shrinkeq}{-2ex}
\begin{equation*}
\left(
\begin{array}{c}
\sqrt{T}B^A_{1} \\
B^A_{2}%
\end{array}%
\right) \Rightarrow -\left( (I_{k_{\beta }};\widetilde{\eta })\Gamma
(I_{k_{\beta }};\widetilde{\eta })^{\prime }+\mathcal{I}_{k_{2}}\Sigma _{u}%
\mathcal{I}_{k_{2}}\right) ^{-1}\mathcal{I}_{k_{2}}\Sigma _{u}\lambda .
\end{equation*}
\end{shrinkeq}

For the ``omitted variable bias,''
\begin{shrinkeq}{-1ex}
\begin{equation*}
\left(
\begin{array}{c}
\sqrt{T}B^{OV}_{1} \\
B^{OV}_{2}%
\end{array}%
\right) =Q_{T}^{-1}\sqrt{T}B^{OV}=\left( \frac{1}{N}\sum_{i=1}^NQ_{T}\widehat{\beta
}_{i}\widehat{\beta }_{i}^{\prime }Q_{T}\right) ^{-1}\frac{1}{N}\sum_{i=1}^NQ_{T}%
\widehat{\beta }_{i}\mu _{i}^{\prime }(\eta _{v,T}-\eta _{T}^{\prime }%
\widetilde{\lambda }).
\end{equation*}
\end{shrinkeq}
Let us consider the following expression:
\begin{shrinkeq}{-1ex}
\begin{equation}
\frac{1}{N}\sum_{i=1}^{N}Q_{T}\widehat{\beta }_{i}\mu _{i}^{\prime }=\frac{1%
}{N}\sum_{i=1}^{N}\left( Q_{T}\beta _{i}+Q_{T}\frac{\eta _{T}\mu _{i}}{\sqrt{%
T}}+Q_{T}u_{i}\right) \mu _{i}^{\prime }.  \label{eq: some eq}
\end{equation}
\end{shrinkeq}%
By Assumption LOADINGS, $\frac{1}{N}\sum_{i=1}^NQ_{T}\beta _{i}\mu _{i}^{\prime
}\rightarrow \Gamma_{\beta \mu }$ and $\frac{1}{N}\sum_{i=1}^N\mu _{i}\mu
_{i}^{\prime }\rightarrow \Gamma_{\mu \mu }$, while $Q_{T}\eta
_{T}/\sqrt{T}\Rightarrow \widetilde{\eta }$. The last term in equation (\ref%
{eq: some eq}) is $o_{P}(1)$ by statement (5) of Lemma S1.
Thus,
\begin{shrinkeq}{-1ex}
\begin{equation*}
\frac{1}{N}\sum_{i=1}^{N}Q_{T}\widehat{\beta }_{i}\mu _{i}^{\prime
}\Rightarrow \Gamma_{\beta \mu }+\widetilde{\eta }\Gamma_{\mu \mu }.
\end{equation*}
\end{shrinkeq}%
We also note that
$
\eta _{v,T}-\eta _{T}^{\prime }\widetilde{\lambda }\Rightarrow \eta
_{v}-\eta ^{\prime }\lambda .
$
This implies validity of the asymptotic statement about $B^{OV}$ contained in
Theorem \ref{theorem: FMB fails}.

For the remaining part, by time averaging equation (\ref{eq: main formulation of dgp}) we get
$
\overline{r}_{i}=\beta _{i}^{\prime }\widetilde{\lambda }+\mu _{i}^{\prime }%
\frac{\eta _{v,T}}{\sqrt{T}}+\overline{e}_{i}.
$
Combining the last equation with equation (\ref{eq: first stage estimate}), we obtain
\begin{shrinkeq}{-2ex}
\begin{equation*}
\overline{r}_{i}=\widehat{\beta }_{i}^{\prime }\widetilde{\lambda }%
-u_{i}^{\prime }\widetilde{\lambda }+\frac{\mu _{i}^{\prime }}{\sqrt{T}}%
(\eta _{v,T}-\eta _{T}^{\prime }\widetilde{\lambda })+\overline{e}_{i}.
\end{equation*}
\end{shrinkeq}%
Thus, we arrive at
\begin{shrinkeq}{-2ex}
\begin{align*}
\widehat{\lambda }_{TP}-\widetilde{\lambda }-B^A-B^{OV}& =\left( \sum_{i=1}^N%
\widehat{\beta }_{i}\widehat{\beta }_{i}^{\prime }\right) ^{-1}\left(
-\sum_{i=1}^N(\widehat{\beta }_{i}-u_{i})u_{i}^{\prime }\widetilde{\lambda }%
+\sum_{i=1}^N\widehat{\beta }_{i}\overline{e}_{i}\right)  \\
& =\left( \sum_{i=1}^N\widehat{\beta }_{i}\widehat{\beta }_{i}^{\prime }\right)
^{-1}\left( -\sum_{i=1}^N(\beta _{i}+\frac{\eta _{T}\mu _{i}}{\sqrt{T}}%
+o_{p}(1))u_{i}^{\prime }\widetilde{\lambda }+\sum_{i=1}^N\widehat{\beta }_{i}%
\overline{e}_{i}\right) ,
\end{align*}
\end{shrinkeq}%
so
\begin{shrinkeq}{-2ex}
\begin{align*}
& \sqrt{NT}Q_{T}^{-1}(\widehat{\lambda }_{TP}-\widetilde{\lambda }-B^A-B^{OV}) \\
& \qquad =\left( \frac{1}{N}\sum_{i=1}^NQ_{T}\widehat{\beta }_{i}\widehat{\beta }%
_{i}^{\prime }Q_{T}\right) ^{-1}\sqrt{\frac{T}{N}}\left(
-\sum_{i=1}^NQ_{T}(\beta _{i}+\frac{\eta _{T}\mu _{i}}{\sqrt{T}}%
)u_{i}^{\prime }\widetilde{\lambda }+\sum_{i=1}^NQ_{T}\widehat{\beta }%
_{i}\overline{e}_{i}\right) .
\end{align*}
\end{shrinkeq}%
Let us prove that the numerator is asymptotically $O_{p}(1)$:
\begin{shrinkeq}{-1ex}
\begin{align}
& \sqrt{\frac{T}{N}}\left( -\sum_{i=1}^NQ_{T}(\beta _{i}+\frac{\eta _{T}\mu _{i}%
}{\sqrt{T}})u_{i}^{\prime }\widetilde{\lambda }+\sum_{i=1}^NQ_{T}\widehat{\beta }%
_{i}\overline{e}_{i}\right)   \notag \\
& \qquad =(I_{k_{\beta }};\widetilde{\eta }_{T})\sqrt{\frac{T}{N}}\sum_{i=1}^N\gamma
_{i}(\overline{e}_{i}-u_{i}^{\prime }\widetilde{\lambda })+\sqrt{\frac{T}{N}}%
\sum_{i=1}^NQ_{T}u_{i}\overline{e}_{i}+O_{p}(1).  \label{eq: one more}
\end{align}
\end{shrinkeq}%
By statement (5) of Lemma S1, we have $\sqrt{%
\frac{T}{N}}\sum_{i}\gamma _{i}\overline{e}_{i}=O_{p}(1)$ and $\sqrt{\frac{T%
}{N}}\sum_{i}\gamma _{i}u_{i}^{\prime }=O_{p}(1)$, which makes the first
summand in equation (\ref{eq: one more}) $O_{p}(1)$. Consider the second
term in equation (\ref{eq: one more}) and recall that $Q_{T}/\sqrt{T}=O(1)$:
\begin{shrinkeq}{-1ex}
\begin{align*}
\sqrt{\frac{T}{N}}\sum_{i=1}^NQ_{T}u_{i}\overline{e}_{i}& =\frac{Q_{T}}{\sqrt{T}}%
\Sigma _{F}^{-1}\frac{1}{\sqrt{N}T}\sum_{i=1}^N\sum_{t=1}^T\sum_{s=1}^T\widetilde{F}%
_{s}e_{is}e_{it} \\
& =\frac{Q_{T}}{\sqrt{T}}\Sigma _{F}^{-1}\frac{1}{\sqrt{N}T}%
\sum_{i=1}^N\sum_{t=1}^T\sum_{s\neq t}\widetilde{F}_{s}e_{is}e_{it}+\frac{Q_{T}}{%
\sqrt{T}}\Sigma _{F}^{-1}\frac{\sqrt{N}}{T}\sum_{t=1}^T\widetilde{F}_{t}S_{t}.
\end{align*}
\end{shrinkeq}%
The first term is $O_{p}(1)$ by statement (4) of Lemma S1,
while the second term is $O_{p}(1)$ by Assumption ERRORS(iii). This ends the
proof of Theorem \ref{theorem: FMB fails}. $\Box $

\textbf{Proof of Corollary \protect\ref{theorem: FMB fails no missing}.}
When we have weak included factors ($k_{2}\geq 1$) but no strong excluded
factors ($k_{v}=0$), the expression for the omitted variables bias ($\mu
_{i}=0$) is exactly equal to zero: $B^{OV}=0$. In this case, $v_{t}=0$ and hence $%
\eta _{T}=0$, as well as $\eta =0$ and $\widetilde{\eta }=0$. That gives the
expression (\ref{eq: ab in no missing case}) for the limit of the
attenuation bias. $\Box $

\textbf{Proof of Corollary \protect\ref{cor: FMB works}.}
If all observed factors are strong, then there is no second component to the
risk premia, i.e., $k_{2}=0$ and $\lambda =\lambda _{1}$. We also have $%
\mathcal{I}_{k_{2}}=0_{k_{F},k_{F}}$ and $\widetilde{\eta }=0$. When applied
to the result of Theorem \ref{theorem: FMB fails}, we obtain that $\sqrt{T}%
B^A\rightarrow ^{p}0$ and
\begin{shrinkeq}{-2ex}
\begin{equation*}
\sqrt{T}B^{OV}\Rightarrow \left( \Gamma_{\beta \beta }\right) ^{-1}%
\Gamma_{\beta \mu }(\eta _{v}-\eta ^{\prime }\lambda ),
\end{equation*}
\end{shrinkeq}%
which is a zero mean gaussian limit. Thus, in this case we have
\begin{shrinkeq}{-2ex}
\begin{equation*}
\sqrt{T}(\widehat{\lambda }-\lambda )=\sqrt{T}(\widetilde{\lambda }-\lambda
)+\sqrt{T}B^{OV}+o_{p}(1).
\end{equation*}
\end{shrinkeq}%
Finally, if in addition to $k_{2}=0$ we also have $k_{v}=0$ (no missing
factor structure), then $B^{OV}=0$ exactly. $\Box $

\textbf{Proof of Theorem \ref{theorem: 4split works}.}
We first discuss the asymptotics of just one IV regression described on step (2), then this argument will be repeated for the other three IV regressions from step (2) of the algorithm. Denote $\tau=\lfloor\frac{T}{4}\rfloor=|T_j|$.

The time-series regression on a sub-sample $j$ gives us that
\begin{shrinkeq}{-1ex}
$$
\widehat\beta^{(j)}_i=\left(\beta_i+u_i^{(j)}+\frac{\eta_{j,T}\mu_i}{\sqrt{\tau}}\right)(1+o_p(1)),
$$
\end{shrinkeq}%
where
$
\eta_{j,T}=\frac{1}{\sqrt{\tau}}\sum_{t\in T_j}\Sigma_F^{-1}\widetilde{F}_t^{(j)}v_t^\prime\Rightarrow\eta_j,
$
$\eta_j$ is random $k_F\times k_v$ matrix with the distribution
$\mathrm{vec}(\eta_j)\sim N(0_{k_Fk_v,1},\Omega_{vF})$, and the $o_p(1)$ term is related to the difference between $\Sigma_F$ and $\frac{1}{\tau}\sum_{t\in T_j}\widetilde{F}_t^{(j)}\widetilde{F}_t^{(j)\prime}$.

On step (2) we run an IV regression of $y_i=\frac{1}{T}\sum_{t=1}^Tr_{it}$ on the regressor
\begin{shrinkeq}{-1ex}
$$x_i^{(1)}=\left(
\begin{array}{c}
  \widehat\beta_i^{(1)} \\
  A_1(\widehat\beta_{i}^{(1)}-\widehat\beta_{i}^{(2)})
\end{array}
\right)=\left(
\begin{array}{c}
  \widehat\beta_i^{(1)} \\
    A_1\frac{\eta_{1,T}-\eta_{2,T}}{\sqrt{\tau}}\mu_i+A_1(u_{i}^{(1)}-u_{i}^{(2)})
\end{array}
\right),
$$
\end{shrinkeq}%
with the instruments
\begin{shrinkeq}{-1ex}
$$
z_i^{(1)}=\left(
\begin{array}{c}
  \widehat\beta_{i}^{(3)} \\
    \widehat\beta_{i}^{(3)}-\widehat\beta_{i}^{(4)}
\end{array}
\right)=
\left(
\begin{array}{c}
  \widehat\beta_{i}^{(3)} \\
  \frac{\eta_{3,T}-\eta_{4,T}}{\sqrt{\tau}}\mu_i+(u_{i}^{(3)}-u_{i}^{(4)})
\end{array}
\right).
$$
\end{shrinkeq}
The main estimation equation can be written in the following way:
\begin{shrinkeq}{-1ex}
\begin{align*}
y_i=&\,\,\frac{1}{T}\sum_{t\in T}r_{it}=\widetilde\lambda^\prime\beta_i+\frac{\eta_{v,T}^\prime}{\sqrt{T}}\mu_i+\overline{e}_i
=\widetilde\lambda^\prime\widehat\beta_i^{(1)}+ \left(\frac{\eta_{v,T}^\prime}{\sqrt{T}}-\widetilde\lambda^{\prime}\frac{\eta_{1,T}}{\sqrt{\tau}}\right)\mu_i +\overline{e}_{i}-\widetilde\lambda^\prime u_i^{(1)}\\
=&\,\,\widetilde\lambda^\prime\widehat\beta_i^{(1)}+a_{1,T}A_1(\widehat\beta_{i}^{(1)}-\widehat\beta_{i}^{(2)})+\overline{e}_{i}- \widetilde\lambda^\prime u_i^{(1)}- a_{1,T}A_1(u_{i}^{(1)}-u_{i}^{(2)}).
\end{align*}
\end{shrinkeq}%
Thus, we can write it as follows:
\begin{shrinkeq}{-2ex}
\begin{align}\label{eq: second stage estimation equation}
y_i=(\widetilde\lambda^\prime,a_{1,T})x_i^{(1)}+\epsilon_i^{(1)}.
\end{align}
\end{shrinkeq}%
Here we use the following notation:
\begin{shrinkeq}{-1ex}
\begin{align*}
a_{1,T}=&\left(\frac{\eta_{v,T}^\prime}{\sqrt{T}}-\widetilde\lambda^\prime \frac{\eta_{1,T}}{\sqrt{\tau}}\right) \left(A_1\frac{\eta_{1,T}-\eta_{2,T}}{\sqrt{\tau}}\right)^{-1}\Rightarrow \left(\frac{\eta_{v}^\prime}{2}-\eta_{1}\right)(A_1(\eta_{1}-\eta_{2}))^{-1},
\end{align*}
\end{shrinkeq}%
and
$
\epsilon_i^{(1)}=\overline{e}_{i}- \widetilde\lambda^\prime u_i^{(1)}- a_{1,T}A_1\left(u_{i}^{(1)}-u_{i}^{(2)}\right).
$
Notice that $a_{1,T}$ is a random $1\times k_v $ matrix that is well defined with probability approaching 1 (as $\eta_{1,T}$ and $\eta_{2,T}$ weakly converge to two independent random gaussian matrices), and $a_{1,T}$ is asymptotically of order $O_p(1)$.

The estimator computed on the step (2) of the four-split algorithm is
\begin{shrinkeq}{-1ex}
\begin{equation*}
\widehat\lambda^{(1)}=(I_{k_F}, 0_{k_F,k_v})\left(X^{(1)\prime}Z^{(1)}(Z^{(1)\prime}Z^{(1)})^{-1}Z^{(1)\prime}X^{(1)}\right)^{-1}X^{(1)\prime}Z^{(1)} (Z^{(1)\prime}Z^{(1)})^{-1}Z^{(1)\prime}Y.
\end{equation*}
\end{shrinkeq}%
Using equation (\ref{eq: second stage estimation equation}) we obtain:
\begin{shrinkeq}{-1ex}
\begin{align}
\widehat\lambda^{(1)}-\widetilde\lambda
= (I_{k_F}, 0_{k_F,k_v})\left(X^{(1)\prime}P_{Z^{(1)}} X^{(1)}\right)^{-1}X^{(1)\prime}P_{Z^{(1)}} \epsilon^{(1)},\label{eq: tsls expression}
\end{align}
\end{shrinkeq}%
where $P_Z$ is a projection matrix onto $Z$.
Let us introduce two normalizing matrices:
\begin{shrinkeq}{-1ex}
$$
Q_x=
    \left(
      \begin{array}{cc}
        Q_T & 0_{k_F,k_v} \\
        0_{k_v,k_F} & \sqrt{T}I_{k_v} \\
      \end{array}
    \right)
    ,
 ~~~~~~~
 Q_z=\left(
      \begin{array}{cc}
        Q_T & 0_{k_F,k_F} \\
        0_{k_F,k_F} & \sqrt{T}I_{k_F} \\
      \end{array}
    \right).
$$
\end{shrinkeq}%
The dimensionality of $Q_x$ is $k\times k$, where $k=k_F+k_v$ is a number of regressors in the second stage regression, while the dimensionality of $Q_z$ is $2k_F\times 2k_F$, where $2k_F$ is the number of instruments. The matrix $Q_T
$ was defined in the proof of Theorem \ref{theorem: FMB fails}. Now,
\begin{shrinkeq}{-1ex}
\begin{align*}
Q_xx_{i}^{(1)}=
\left(\widetilde{A}_{1,T}\gamma_i+\left(
                \begin{array}{cc}
                  Q_T & 0_{k_F,k_F} \\
                  \sqrt{T}A_1 & -\sqrt{T}A_1 \\
                \end{array}
              \right)
              \left(
                \begin{array}{c}
                  u_i^{(1)} \\
                  u_i^{(2)} \\
                \end{array}
              \right)
\right),
\end{align*}
\end{shrinkeq}%
where
\begin{shrinkeq}{-1ex}
\begin{align*}
\widetilde{A}_{1,T}=\left(
                      \begin{array}{cc}
                        I_{k_F} & Q_T\frac{\eta_{1,T}}{\sqrt{\tau}}\\
                        0_{k_v,k_F} & 2A_1(\eta_{1,T}-\eta_{2,T}) \\
                      \end{array}
                    \right)&\Rightarrow\left(
                      \begin{array}{cc}
                        I_{k_F} & 2\mathcal{I}_{k_2}\eta_1 \\
                        0_{k_v,k_F} & 2A_1(\eta_{1}-\eta_{2}) \\
                      \end{array}
                    \right)= \widetilde{A}_{1},\\
\frac{1}{\sqrt{T}}\left(
                \begin{array}{cc}
                  Q_T & 0_{k_F,k_F} \\
                  \sqrt{T}A_1 & -\sqrt{T}A_1 \\
                \end{array}
              \right)&\rightarrow
              \left(
                \begin{array}{cc}
                  \mathcal{I}_{k_2} & 0_{k_F,k_F} \\
                  A_1 & -A_1 \\
                \end{array}
              \right).
\end{align*}
\end{shrinkeq}%
Here $\mathcal{I}_{k_2}$ is a $k_F\times k_F$ matrix which was introduced in Theorem \ref{theorem: FMB fails}.
We also have
\begin{shrinkeq}{-1ex}
\begin{align*}
Q_zz_{i}^{(1)}=
\left(A^*_{1,T}\gamma_i+
\left(
                \begin{array}{cc}
                  Q_T/\sqrt{T}& 0_{k_F,k_F} \\
                  I_{k_F} & -I_{k_F} \\
                \end{array}
              \right)\sqrt{T}
              \left(
                \begin{array}{c}
                  u_i^{(3)} \\
                  u_i^{(4)} \\
                \end{array}
              \right)
\right),
\end{align*}
\end{shrinkeq}%
where
\begin{shrinkeq}{-1ex}
\begin{align*}
A^*_{1,T}=\left(
                      \begin{array}{cc}
                        I_{k_F} & Q_T\frac{\eta_{3,T}}{\sqrt{\tau}} \\
                        0_{k_F,k_F} & 2(\eta_{3,T}-\eta_{4,T}) \\
                      \end{array}
                    \right)&\Rightarrow \left(
                      \begin{array}{cc}
                        I_{k_F} & 2\mathcal{I}_{k_2}\eta_3 \\
                        0_{k_F,k_F} & 2(\eta_{3}-\eta_{4}) \\
                      \end{array}
                    \right)=A^*_{1},
\\
\left(
                \begin{array}{cc}
                  Q_T/\sqrt{T}& 0_{k_F,k_F} \\
                  I_{k_F} & -I_{k_F} \\
                \end{array}
              \right)&\rightarrow
              \left(
                \begin{array}{cc}
                  \mathcal{I}_{k_2} & 0_{k_F,k_F} \\
                  I_{k_F} & -I_{k_F} \\
                \end{array}
              \right).
\end{align*}
\end{shrinkeq}%
Statements (1) and (5) of Lemma S1 imply that
\begin{shrinkeq}{-2ex}
\begin{align}\label{eq: basic convergence1}
\frac{T}{\sqrt{N}}\sum_{i=1}^Nu_i^{(j)}u_i^{(j^*)\prime}&=O_p(1) ~~~~\mbox{for}~~~j\neq j^*,
\\
\label{eq: basic convergence2}
\sqrt{\frac{T}{N}}\sum_{i=1}^N(\gamma_i^\prime,1)^\prime u_i^{(j^*)\prime}&=O_p(1) .
\end{align}
\end{shrinkeq}%
This  together with Assumption LOADINGS gives us that
\begin{shrinkeq}{-1ex}
\begin{align}\label{eq: zx limit}
\frac{1}{N}\sum_{i=1}^NQ_xx_{i}^{(1)}z_i^{(1)\prime}Q_z\Rightarrow \widetilde{A}_1\Gamma A^{*\prime}_1.
\end{align}
\end{shrinkeq}%
By Assumption LOADINGS, $\Gamma$ is full rank, while $\widetilde{A}_1$ and $A^{*\prime}_1$ are full rank with probability 1.
Statements (3) and (4) of Lemma S1 imply that
\begin{shrinkeq}{-1ex}
\begin{align}\label{eq: basic convergence1}
\frac{\tau}{N}\sum_{i=1}^Nu_i^{(j)}u_i^{(j)\prime}\to^p \Sigma_u.
\end{align}
\end{shrinkeq}%
Thus, we obtain
\begin{shrinkeq}{-2ex}
\begin{align}\notag
\frac{1}{N}\sum_{i=1}^NQ_zz_{i}^{(1)}z_i^{(1)\prime}Q_z &\Rightarrow A^*_1\Gamma A^{*\prime}_1+4\left(
                \begin{array}{cc}
                  \mathcal{I}_{k_2} & 0_{k_F,k_F} \\
                  I_{k_F} & -I_{k_F} \\
                \end{array}
              \right)
              \left(
                \begin{array}{cc}
                  \Sigma_u & 0_{k_F,k_F} \\
                  0_{k_F,k_F} & \Sigma_u \\
                \end{array}
              \right)
              \left(
                \begin{array}{cc}
                  \mathcal{I}_{k_2} & I_{k_F} \\
                   0_{k_F,k_F}& -I_{k_F} \\
                \end{array}
              \right)\\& = A^*_1\Gamma A^{*\prime}_1+4\left(
                \begin{array}{cc}
                  \mathcal{I}_{k_2}\Sigma_u\mathcal{I}_{k_2} &  \mathcal{I}_{k_2}\Sigma_u\\
                  \Sigma_u\mathcal{I}_{k_2}& 2\Sigma_u \\
                \end{array}
              \right).\label{eq: xxlimit}
\end{align}
\end{shrinkeq}%
Let us now show that
\begin{shrinkeq}{-1ex}
\begin{align}\label{eq: error term}
\sqrt{\frac{T}{N}}\sum_{i=1}^NQ_zz_i^{(1)}\epsilon_i^{(1)}=O_p(1).
\end{align}
\end{shrinkeq}%
We have
$
\epsilon_i^{(1)}=\overline{e}_{i}- \widetilde\lambda^\prime u_i^{(1)}- a_{1,T}A_1(u_{i}^{(1)}-u_{i}^{(2)}).
$
The sum in (\ref{eq: error term}) contains summands of the form
$
\sqrt{\frac{T}{N}}\sum_{i=1}^N\gamma_i(\overline{e}_i,u_i^{(j)})$,
$\frac{T}{\sqrt{N}}\sum_{i=1}^N\overline{e}_iu_i^{(j)}$ and $\frac{T}{\sqrt{N}}\sum_{i=1}^Nu_i^{(j^*)\prime}u_i^{(j)}$. All three types of summands are $O_p(1)$ due to statements (5), (2) and (1) of Lemma S1,  correspondingly.
Putting equations (\ref{eq: zx limit}) and (\ref{eq: xxlimit}) together, we obtain
\begin{shrinkeq}{-1ex}
\begin{align*}
NQ_x^{-1}\Theta_{N,T,1}Q_z^{-1}
=& \left(\frac{Q_xX^{(1)\prime} Z^{(1)}Q_z}{N}\left(\frac{Q_zZ^{(1)\prime}Z^{(1)}Q_z}{N}\right)^{-1} \frac{Q_zZ^{(1)\prime}X^{(1)}Q_x}{N}\right)^{-1} \\ &\quad\cdot \frac{Q_xX^{(1)\prime}Z^{(1)}Q_z}{N} \left(\frac{Q_zZ^{(1)\prime}Z^{(1)}Q_z}{N}\right)^{-1}= O_p(1).
\end{align*}
\end{shrinkeq}%
Putting everything together, we have:
\begin{shrinkeq}{-2ex}
$$\sqrt{NT}Q^{-1}_T(\widehat\lambda^{(1)}-\widetilde\lambda)=(I_{k_F}, 0_{k_F,k_v})NQ_x^{-1}\Theta_{N,T,1}Q_z^{-1}\sqrt{\frac{T}{N}}\sum_{i=1}^NQ_zz_{i}^{(1)} \epsilon^{(1)}_i=O_p(1).
$$
\end{shrinkeq}

Because
$\sqrt{NT}Q^{-1}_T=\left(
                   \begin{array}{cc}
                     \sqrt{NT}I_{k_1} & 0_{k_1,k_2} \\
                     0_{k_2,k_1} & \sqrt{N}I_{k_2} \\
                   \end{array}
                 \right),
$ we obtain different rates of estimation of the risk premia $\lambda_1$ and $\lambda_2$ on the strong and weak observed factors.
We have
$
\sqrt{NT}(\widehat\lambda^{(1)}_1-\widetilde\lambda_1)=O_p(1),
$
while
$
\sqrt{N}(\widehat\lambda^{(1)}_2-\widetilde\lambda_2)=O_p(1).
$
Thus,
\begin{shrinkeq}{-1ex}
\begin{align*}
\sqrt{T}(\widehat\lambda^{(1)}_1-\lambda_1)=\sqrt{T}(\widetilde\lambda_1-\lambda_1)+ \sqrt{T}(\widehat\lambda^{(1)}_1-\widetilde\lambda_1) =\sqrt{T}(\widetilde{\lambda}_{1}-\lambda_1)+O_p(1/\sqrt{N})\Rightarrow N(0,\Omega_F).
\end{align*}
\end{shrinkeq}%
As for the estimator of the risk premia on the weak factors,
\begin{shrinkeq}{-1ex}
$$
\widehat\lambda^{(1)}_2-\lambda_2=(\widetilde\lambda_2-\lambda_2)+ (\widehat\lambda^{(1)}_2-\widetilde\lambda_2) =O_p(1/\sqrt{T})+O_p(1/\sqrt{N})=O_p(1/\sqrt{\min\{N,T\}}).
$$
\end{shrinkeq}%
We have proved the statement of Theorem \ref{theorem: 4split works} for an estimator obtained on step (2) of the algorithm, but the same line of reasoning applies to $\widehat\lambda^{(2)}, \widehat\lambda^{(3)}, \widehat\lambda^{(4)}$ and their average. This finishes the proof of Theorem \ref{theorem: 4split works}.
$\Box$

\textbf{Proof of Theorem \ref{theorem: asymptotics of 4 split}.}
Following the steps of the proof of Theorem \ref{theorem: 4split works} we get the following two statements:
\begin{shrinkeq}{-2ex}
\begin{align}\label{eq: general xz}
\frac{1}{N}\sum_{i=1}^NQ_xx_{i}^{(j)}z_i^{(j)\prime}Q_z\Rightarrow & \, \widetilde{A}_j\mathbf{\Gamma}A^{*\prime}_j,
\\ \label{eq: general xx}
\frac{1}{N}\sum_{i=1}^NQ_zz_{i}^{(j)}z_i^{(j)\prime}Q_z  \Rightarrow & \, A^*_j\mathbf{\Gamma}A^{*\prime}_j+4\left(
                \begin{array}{cc}
                  \mathcal{I}_{k_2}\Sigma_u\mathcal{I}_{k_2} &  \mathcal{I}_{k_2}\Sigma_u\\
                  \Sigma_u\mathcal{I}_{k_2}& 2\Sigma_u \\
                \end{array}
              \right),
\end{align}
\end{shrinkeq}%
where $A^*_j$ and $\widetilde{A}_j$ are random matrices that are deterministic functions of random vectors $(\eta_1,...,\eta_4)$.
Indeed, let us adopt the following notation. Let $j_1,...,j_4$ be the circular indexes used for computing $\widehat{\lambda}^{(j)}$.
In particular, the estimate $\widehat{\lambda}^{(j)}$ is computed from the IV regression with the regressors  $x_{i}^{(j)}=\left( \widehat{\beta }_{i}^{(j_1)\prime },(%
\widehat{\beta }_{i}^{(j_1)}-\widehat{\beta }_{i}^{(j_2)})^{\prime
}A_{j}^{\prime }\right) ^{\prime }$ and the instruments $z_{i}^{(j)}=\left(
\widehat{\beta }_{i}^{(j_3)\prime },(\widehat{\beta }_{i}^{(j_3)}-\widehat{\beta
}_{i}^{(j_4)})^{\prime }\right) ^{\prime }$. Then, similarly to the proof of Theorem \ref{theorem: 4split works}, we obtain:
\begin{shrinkeq}{-1ex}
\begin{align*}
A^*_{j}=\left(
                      \begin{array}{cc}
                        I_{k_F} & 2\mathcal{I}_{k_2}\eta_{j_3} \\
                        0_{k_F,k_F} & 2(\eta_{j_3}-\eta_{j_4}) \\
                      \end{array}
                    \right), ~~~~~
                    \widetilde{A}_{j}=\left(
                      \begin{array}{cc}
                        I_{k_F} & 2\mathcal{I}_{k_2}\eta_{j_1} \\
                        0_{k_v,k_F} & 2A_j(\eta_{j_1}-\eta_{j_2}) \\
                      \end{array}
                    \right).
\end{align*}
\end{shrinkeq}%
So,
\begin{shrinkeq}{-2ex}
\begin{align*}
&NQ_x^{-1}\left(X^{(j)\prime}Z^{(j)}(Z^{(j)\prime}Z^{(j)})^{-1}Z^{(j)\prime}X^{(j)}\right)^{-1}X^{(j)\prime}Z^{(j)} (Z^{(j)\prime}Z^{(j)})^{-1}Q_z^{-1}\Rightarrow\Theta_j.
\end{align*}
\end{shrinkeq}%
The limit  $\Theta_j$ in the last expression is a known deterministic function of random vectors $(\eta_1,...,\eta_4)$, which can be explicitly written  in terms of $A^*_j$ and $\widetilde{A}_j$.

We have the following expression for the estimates obtained on steps (2) and (3) of the four-split algorithm:
\begin{shrinkeq}{-2ex}
$$\sqrt{NT}Q_T^{-1}(\widehat\lambda^{(j)}-\widetilde\lambda)=(I_{k_F}, 0_{k_F,k_v})NQ_x^{-1}\Theta_{N,T,j}Q_z^{-1}\sqrt{\frac{T}{N}}\sum_{i=1}^NQ_zz_{i}^{(j)} \epsilon^{(j)}_i,
$$
\end{shrinkeq}%
where
$
\epsilon_i^{(j)}=\overline{e}_{i}- \widetilde\lambda^\prime u_i^{(j_1)}- a_{j,T}A_j(u_{i}^{(j_1)}-u_{i}^{(j_2)}),$ and
\begin{shrinkeq}{-1ex}
\begin{align*}
Q_zz_{i}^{(j)}=&
A^*_{j,T}\gamma_i+
\left(
                \begin{array}{cc}
                  Q_T/\sqrt{T}& 0_{k_F,k_F} \\
                  I_{k_F} & -I_{k_F} \\
                \end{array}
              \right)\sqrt{T}
              \left(
                \begin{array}{c}
                  u_i^{(j_3)} \\
                  u_i^{(j_4)} \\
                \end{array}
              \right).
\end{align*}
\end{shrinkeq}%
Consider the following term which can be rewritten in terms of $\xi_i$ from Assumption GAUSSIANITY:
\begin{shrinkeq}{-2ex}
\begin{align*}
&\sqrt{\frac{T}{N}}\sum_{i=1}^NQ_zz_{i}^{(j)} \epsilon^{(j)}_i=A^*_{j,T}\left(\sqrt{\frac{T}{N}}\sum_{i=1}^N\gamma_i
\left(
  \begin{array}{c}
    \overline{e}_i \\
    u_{i}^{(j_1)} \\
    u_i^{(j_2)} \\
  \end{array}
\right)^\prime\right)\left(
                \begin{array}{c}
                  1 \\
                  -\widetilde\lambda-A_j'a_{j,T}' \\
                  A_j'a_{j,T}' \\
                \end{array}
              \right)\\
              & \qquad +\left(
                \begin{array}{cc}
                  Q_T/\sqrt{T}& 0_{k_F,k_F} \\
                  I_{k_F} & -I_{k_F} \\
                \end{array}
              \right)\left(\frac{T}{\sqrt{N}}\sum_{i=1}^N
              \left(
                \begin{array}{c}
                  u_i^{(j_3)} \\
                  u_i^{(j_4)} \\
                \end{array}
              \right)\left(
  \begin{array}{c}
    \overline{e}_i \\
    u_{i}^{(j_1)} \\
    u_i^{(j_2)} \\
  \end{array}
\right)^\prime\right)\left(
                \begin{array}{c}
                  1 \\
                  -\widetilde\lambda-A_j'a_{j,T}' \\
                  A_j'a_{j,T}' \\
                \end{array}
              \right)\\&\quad =\mathcal{A}_{j,T}\frac{1}{\sqrt{N}}\sum_{i=1}^N\xi_i,
\end{align*}
\end{shrinkeq}%
where $\mathcal{A}_{j,T}$ is a $k_z\times k_\xi$ matrix which is a deterministic function of $A^*_{j,T}, A_j, a_{j,T},\widetilde\lambda$. The exact expression for $\mathcal{A}_{j,T}$ is  obvious though too complicated to write down. We have discussed before the convergence of all terms separately, which implies that $\mathcal{A}_{j,T}\Rightarrow \mathcal{A}_{j}$, where the limit is a deterministic function of $(\eta_1,...,\eta_4)$.

Given Assumption GAUSSIANITY, we  have
$
\sqrt{\frac{T}{N}}\sum_{i=1}^NQ_zz_{i}^{(j)} \epsilon^{(j)}_i\Rightarrow\mathcal{A}_j\xi.
$
Following step (4) of the four-split algorithm, we can put all pieces together:
\begin{shrinkeq}{-1ex}
\begin{align}\label{eq: mixed normal}
\sqrt{NT}Q^{-1}(\widehat\lambda_{4S}-\widetilde\lambda)\Rightarrow(I_{k_F}, 0_{k_F,k_v})\left(\frac{1}{4}\sum_{j=1}^4\Theta_j\mathcal{A}_j\right)\xi.
\end{align}
\end{shrinkeq}%
As we can see, the four-split estimator is asymptotically mixed gaussian; that is, the limit distribution conditionally on $\eta_1,...,\eta_4$ (which is independent of $\xi$ due to Assumption ERRORS) is gaussian with mean zero and the variance depending on $\eta_1,...,\eta_4$.

Denote $
\widehat\Sigma_{IV}=\frac{1}{N}R^{\prime }G^{-1}\widehat{\Sigma }_{0}G^{-1}R.
$
We show below that $\widehat\Sigma_{IV}$ has the following asymptotic distribution:
\begin{shrinkeq}{-1ex}
\begin{align}\label{eq: asymptotics of variance}
NTQ^{-1}_T\widehat\Sigma_{IV}Q^{-1}_T\Rightarrow (I_{k_F}, 0_{k_F,k_v})\left(\frac{1}{4}\sum_{j=1}^4\Theta_j\mathcal{A}_j\right)\Sigma_\xi \left(\frac{1}{4}\sum_{j=1}^4\Theta_j\mathcal{A}_j\right)^\prime (I_{k_F}, 0_{k_F,k_v})^\prime.
\end{align}
\end{shrinkeq}%
Statement (\ref{eq: asymptotics of variance})  implies the statement of Theorem \ref{theorem: asymptotics of 4 split}. Indeed,
equations (\ref{eq: mixed normal}) and (\ref{eq: asymptotics of variance}) imply that
$\widehat\Sigma_{IV}^{-1/2}(\widehat\lambda_{4S}-\widetilde\lambda)\Rightarrow N(0,I_k),$
where the limiting gaussian vector  is  independent of the limiting gaussian vector in the following convergence:
\begin{shrinkeq}{-1ex}
$$
\sqrt{T}\Omega_F^{-1/2}(\widetilde\lambda-\lambda)\Rightarrow N(0,I_k).
$$
\end{shrinkeq}%
The expression $\widehat\Sigma_{4S}^{-1/2}(\widehat{\lambda}_{4S}-\lambda)$ is the weighted sum of the expressions staying on the left-hand-side of the last two convergence with weights asymptotically independent from both limiting $N(0,I_k)$. This leads to the validity of the statement of Theorem \ref{theorem: asymptotics of 4 split}.

To prove the validity of statement (\ref{eq: asymptotics of variance}) we notice that
$\sqrt{T}Q_zz_i^{(j)}\epsilon_i^{(j)}=\mathcal{A}_{j,T}\xi_i$. Thus,
\begin{shrinkeq}{-2ex}
\begin{align}\notag
\frac{T}{N}\sum_{i=1}^{N}\left(
\begin{array}{c}
Q_z\widetilde{z}_{i}^{(1)}\epsilon_{i}^{(1)} \\
... \\
Q_z\widetilde{z}_{i}^{(4)}\epsilon_{i}^{(4)}%
\end{array}%
\right) \left(
\begin{array}{c}
Q_z\widetilde{z}_{i}^{(1)}\epsilon_{i}^{(1)} \\
... \\
Q_z\widetilde{z}_{i}^{(4)}\epsilon_{i}^{(4)}%
\end{array}%
\right) ^{\prime }= & \left(
\begin{array}{c}
\mathcal{A}_{1,T}\\
... \\
\mathcal{A}_{4,T}%
\end{array}%
\right) \frac{1}{N}\sum_{i=1}^{N}\xi_i\xi_i^\prime\left(
\begin{array}{c}
\mathcal{A}_{1,T}\\
... \\
\mathcal{A}_{4,T}%
\end{array}%
\right) ^{\prime }\\
\Rightarrow & \left(
\begin{array}{c}
\mathcal{A}_{1}\\
... \\
\mathcal{A}_{4}%
\end{array}%
\right) \Sigma_{\xi}\left(
\begin{array}{c}
\mathcal{A}_{1}\\
... \\
\mathcal{A}_{4}%
\end{array}%
\right) ^{\prime }
.\label{eq: infeasible}
\end{align}
\end{shrinkeq}%
Let us consider an infeasible variance estimator $\widetilde\Sigma_{IV}$ which is constructed in the same way as $\widehat\Sigma_{IV}$ but uses $\epsilon_{i}^{(j)}$ in place of $\widehat{\epsilon}_{i}^{(j)}$.
That is, denote
\begin{shrinkeq}{-1ex}
\begin{equation*}
\widetilde{\Sigma }_{0}=\frac{1}{N}\sum_{i=1}^{N}\left(
\begin{array}{c}
\widetilde{z}_{i}^{(1)}\epsilon_{i}^{(1)} \\
... \\
\widetilde{z}_{i}^{(4)}\epsilon_{i}^{(4)}%
\end{array}%
\right) \left(
\begin{array}{c}
\widetilde{z}_{i}^{(1)}\epsilon_{i}^{(1)} \\
... \\
\widetilde{z}_{i}^{(4)}\epsilon_{i}^{(4)}%
\end{array}%
\right) ^{\prime },
\end{equation*}
\end{shrinkeq}%
and consider
$\widetilde\Sigma_{IV}=\frac{1}{N}R^{\prime }G^{-1}\widetilde{\Sigma }_{0}G^{-1}R.$
By putting together (\ref{eq: general xz}), (\ref{eq: general xx}) and (\ref{eq: infeasible}) we obtain
\begin{shrinkeq}{-1ex}
\begin{align*}
NTQ^{-1}_T\widetilde\Sigma_{IV}Q^{-1}_T\Rightarrow (I_{k_F}, 0_{k_F,k_v})\left(\frac{1}{4}\sum_{j=1}^4\Theta_j\mathcal{A}_j\right)\Sigma_\xi \left(\frac{1}{4}\sum_{j=1}^4\Theta_j\mathcal{A}_j\right)^\prime (I_{k_F}, 0_{k_F,k_v})^\prime.
\end{align*}
\end{shrinkeq}%
The only thing left to show is that the difference between $\widehat\Sigma_{IV}$ and $\widetilde{\Sigma}_{IV}$ is asymptotically negligible. In particular, we will show for any $j$ and $j^*$,
\begin{shrinkeq}{-1ex}
\begin{align}\label{eq: residulas vs errors}
\frac{T}{N}\sum_{i=1}^NQ_zz_{i}^{(j)}z_{i}^{(j^*)\prime}Q_z \left(\epsilon^{(j)}_i\epsilon^{(j^*)}_i-\widehat\epsilon^{(j)}_i\widehat\epsilon^{(j^*)}_i\right) \to^p 0,
\end{align}
\end{shrinkeq}%
where $\widehat\epsilon^{(j)}_i$ are the residuals from the $(j)^{th}$ IV regression. Indeed, this last statement implies that $
\widehat\Sigma_{IV}=\widetilde\Sigma_{IV}(1+o_p(1)),$
and usage of residuals in place of true errors does not have an asymptotic effect on estimation of variance.

In order to prove (\ref{eq: residulas vs errors}) we write down an equation analogous to equation (\ref{eq: second stage estimation equation}):
\begin{shrinkeq}{-1ex}
\begin{align*}
y_i=(\widetilde\lambda^\prime,a_{j,T})x_i^{(j)}+\epsilon_i^{(j)}=\theta_j^\prime x_i^{(j)}+\epsilon_i^{(j)}.
\end{align*}
\end{shrinkeq}%
From the proof of Theorem \ref{theorem: 4split works} we have that
$\sqrt{NT}Q_{x}^{-1}(\widehat\theta_j-\theta_j)=O_p(1),$
where $\widehat\theta_j$ is the IV estimator obtained on Steps (2) for $j=1$ or on Step (3) for $j=2,3$ or $4$. The residuals for this regression are
\begin{shrinkeq}{-2ex}
$$
\widehat\epsilon^{(j)}_i=y_i-\widehat\theta_j^\prime x_i^{(j)}=\epsilon_i^{(j)} -(\widehat\theta_j-\theta_j)^\prime x_i^{(j)}=\epsilon_i^{(j)} -(Q_x^{-1}(\widehat\theta_j-\theta_j))^\prime Q_xx_i^{(j)}.
$$
\end{shrinkeq}%
The left hand expression of (\ref{eq: residulas vs errors}) is equal to
\begin{shrinkeq}{-1ex}
\begin{align*}
\frac{T}{N}\sum_{i=1}^NQ_zz_{i}^{(j)}z_{i}^{(j^*)\prime}Q_z \left(\epsilon_i^{(j)}(\widehat\theta_{j^*}-\theta_{j^*})^\prime x_i^{(j^*)}+ \epsilon_i^{(j^*)}(\widehat\theta_{j}-\theta_{j})^\prime x_i^{(j)}- (\widehat\theta_{j^*}-\theta_{j^*})^\prime x_i^{(j^*)}(\widehat\theta_{j}-\theta_{j})^\prime x_i^{(j)}\right).
\end{align*}
\end{shrinkeq}%
This expression contains three sums. We can show that each of them is asymptotically negligible. For example, consider the first of the three sums:
\begin{shrinkeq}{-1ex}
\begin{align*}
\frac{1}{N^{3/2}}\sum_{i=1}^N\left(\sqrt{T}Q_zz_{i}^{(j)}\epsilon_i^{(j)}\right)\left(Q_zz_{i}^{(j^*)} \right)^{\prime}\left(Q_xx_{i}^{(j^*)} \right)^{\prime} \left(\sqrt{NT}Q_x^{-1}(\widehat\theta_{j^*}-\theta_{j^*})\right)\\
=\frac{1}{N^{3/2}}\sum_{i=1}^N\mathcal{A}_{j,T}\xi_i \left(Q_zz_{i}^{(j^*)} \right)^{\prime}\left(Q_xx_{i}^{(j^*)} \right)^{\prime}\left(\sqrt{NT}Q_x^{-1}(\widehat\theta_{j^*}-\theta_{j^*})\right).
\end{align*}
\end{shrinkeq}%
Note that $\sqrt{NT}Q_x^{-1}(\widehat\theta_{j^*}-\theta_{j^*})=O_p(1).$
As before, $Q_zz_i^{(j)}=O_p(1)\gamma_i+O_p(1)\sqrt{T}(u_i^{(j_3)},u_i^{(j_4)})^\prime$, while $Q_xx_i^{(j)}=O_p(1)\gamma_i+O_p(1)\sqrt{T}(u_i^{(j_1)},u_i^{(j_2)})$, where all the mentioned $O_p(1)$ terms are not indexed by $i$. Thus,
\begin{shrinkeq}{-1ex}
$$
\frac{1}{N^{3/2}}\sum_{i=1}^N\mathcal{A}_{j,T}\xi_i \left(Q_zz_{i}^{(j^*)} \right)^{\prime}\left(Q_xx_{i}^{(j^*)} \right)^{\prime}
=O_p(1)\frac{1}{N^{3/2}}\sum_{i=1}^N\xi_i\xi_i^\prime+O_p(1)\frac{1}{N^{3/2}}\sum_{i=1}^N\xi_i\otimes (\gamma_i\gamma_i^\prime).
$$
\end{shrinkeq}%
By Assumption GAUSSIANITY,  $\frac{1}{N^{3/2}}\sum_{i=1}^N\xi_i\xi_i^\prime\to^p 0$ and thus
\begin{shrinkeq}{-1ex}
$$
\frac{1}{N^{3/2}}\left\|\sum_{i=1}^N\xi_i\otimes (\gamma_i\gamma_i^\prime)\right\|\leq \frac{1}{N^{3/2}} \sqrt{\sum_{i=1}^N\|\xi_i\|^2}\sqrt{\sum_{i=1}^N\|\gamma_i\|^4}\to^p0.
$$
\end{shrinkeq}%
This gives the asymptotic negligibility of the first sum; the negligibility of the other two sums is proved in a similar manner. This ends the proof of Theorem \ref{theorem: asymptotics of 4 split}. $\Box$

\end{document}